\documentclass[%
reprint,
superscriptaddress,
preprintnumbers,
nofootinbib,
amsmath,amssymb,
aps,
prd,
]{revtex4-2}

\usepackage{enumitem}
\usepackage{graphicx}
\usepackage{dcolumn}
\usepackage{bm}
\usepackage[colorlinks, linkcolor=blue, pdfborder={0 0 0}, breaklinks=true]{hyperref}

\usepackage{xcolor}
\usepackage{soul}

\newcommand{\mpch}{\>h^{-1}{\rm {Mpc}}}
\newcommand{\hmpc}{\>h~\mathrm{Mpc}^{-1}}

\begin{document}



\title{Deep learning with hybrid frequency differencing and principal component analysis for 21-cm foreground and beam mitigation}

\author{Zitong Wang}
\affiliation{School of Aerospace Science and Technology, Xidian University, Xi'an 710126, China}

\author{Feng Shi}
\altaffiliation{fshi@xidian.edu.cn}
\affiliation{School of Aerospace Science and Technology, Xidian University, Xi'an 710126, China}

\author{Le Zhang}
\affiliation{School of Physics and Astronomy, Sun Yat-sen University, 2 Daxue Road, Tangjia, Zhuhai, 519082, People's Republic of China}
\affiliation{CSST Science Center for the Guangdong-Hong Kong-Macau Greater Bay Area, Zhuhai 519082, People's Republic of China}

\author{Yanming Liu}
\affiliation{School of Aerospace Science and Technology, Xidian University, Xi'an 710126, China}

\author{Xiaoping Li}
\affiliation{School of Aerospace Science and Technology, Xidian University, Xi'an 710126, China}

\author{Shulei Ni}
\affiliation{Research Center for Astronomical Computing,  Zhejiang Laboratory, Hangzhou 311121, China}

\author{Ming Jiang}
\affiliation{National Key Laboratory of Radar Signal Processing, Xidian University, Xi'an 710126, People's Republic of China}

\author{Xiaofan Ma}
\affiliation{School of Aerospace Science and Technology, Xidian University, Xi'an 710126, China}

\date{\today}

\begin{abstract}
Twenty-one-centimeter intensity mapping is a powerful probe of the large-scale distribution of neutral hydrogen (HI) and cosmological observables such as baryon acoustic oscillations. A major challenge is contamination from bright foregrounds and frequency-dependent beam effects, which can lead to signal loss in traditional methods such as principal component analysis (PCA). We develop a hybrid approach that trains a U-shaped convolutional neural network (UNet) on two input channels derived from frequency differencing (FD) and PCA cleaning, enabling it to exploit their complementary behavior across different scales. This two-channel strategy achieves improved performance, maintaining the cross-correlation power spectrum close to unity on large scales under a cosine beam and improving by 5\%–8\% relative to either FD- or PCA-based UNet alone. We further show that the method can robustly recover the HI signal even when the beam model is imperfect and differs between training and testing, with the large-scale cross-correlation remaining close to unity within the $1\sigma$ level. These results demonstrate that the proposed approach provides a robust framework for HI signal reconstruction under realistic observational conditions.

\end{abstract}

\maketitle

\section{Introduction} \label{sec:intro}

The cosmic 21-cm line of neutral hydrogen (HI) has emerged as a powerful tracer of large-scale structure in the postreionization Universe. In particular, 21-cm intensity mapping (IM) has been proposed as an efficient method to map the three-dimensional distribution of HI without resolving individual galaxies \citep{2001JApA...22...21B,2004MNRAS.355.1339B,2009astro2010S.234P,2009MNRAS.397.1926W}. By measuring the aggregate 21-cm brightness temperature fluctuations across low angular resolution voxels, IM surveys can rapidly cover enormous cosmological volumes. This opens a promising avenue to probe key cosmological observables at low redshifts, notably baryon acoustic oscillations (BAOs) as a standard ruler for cosmic expansion, redshift-space distortions (RSDs) as a probe of structure growth, and the bias and abundance of HI as tracers of the underlying matter field. Low-redshift 21-cm surveys offer a novel and complementary probe of cosmology (e.g., \cite{2008PhRvL.100i1303C,2010ApJ...721..164S, 2020PASP..132f2001L, 2023SCPMA..6670413W}), analogous to optical galaxy redshift surveys but with the potential for far greater volume and efficiency.

Over the past decade, a large number of interferometric and single-dish radio telescopes have been conducting, and will continue to carry out, 21-cm intensity mapping observations, including CHIME (Canadian Hydrogen Intensity Mapping Experiment) \citep{2023ApJ...947...16A,2022ApJS..261...29C}, Tianlai \citep{2011SSPMA..41.1358C,2015ApJ...798...40X,2021MNRAS.506.3455W,2022MNRAS.517.4637P}, BINGO (baryon acoustic oscillations in neutral gas observations) \citep{2013MNRAS.434.1239B,2022BINGO}, MeerKAT (the South African Square Kilometre Array Pathfinder) \citep{2016mks..confE..32S,2021MNRAS.505.3698W}, HIRAX (the Hydrogen Intensity and Real-Time Analysis Experiment) \citep{2016SPIE.9906E..5XN}, and the SKA (Square Kilometre Array) \citep{2015aska.confE..19S}.
A series of pioneering experiments have demonstrated the viability of 21-cm intensity mapping at low redshift. The first detections of cosmological 21-cm fluctuations were obtained through cross-correlation of Green Bank Telescope intensity maps with galaxy redshift surveys \cite{2010Natur.466..463C, 2013ApJ...763L..20M, 2022MNRAS.510.3495W}, followed by similar measurements using the Parkes telescope\cite{2018MNRAS.476.3382A}. The CHIME achieved the first interferometric detection of the 21-cm signal, reporting a statistically significant cross-correlation with eBOSS tracers\cite{2023ApJ...947...16A}. More recently, a pilot intensity mapping survey with the MeerKAT array has produced the first high-significance detection of postreionization HI signal, measured via cross-correlation with the WiggleZ galaxy survey at $0.4 < z < 0.459$ \cite{2023MNRAS.518.6262C}. This represents the first practical demonstration of the multidish autocorrelation intensity mapping technique for cosmology and an important step toward a direct measurement of the HI autopower spectrum, with deep MeerKAT observations already beginning to place upper limits on this quantity \cite{2025MNRAS.541..476M}. These results mark the transition of low-redshift 21-cm cosmology from proof-of-concept studies toward statistically robust detections, establishing the groundwork for future measurements of cosmology. 

However, directly recovering the cosmological 21-cm signal from intensity maps remains challenging because astrophysical foregrounds, dominated by Galactic synchrotron, free-free emission and extragalactic point sources, are about 4 orders of magnitude brighter than the HI signal~\cite{2002ApJ...564..576D,2003MNRAS.346..871O,2005ApJ...625..575S,2008MNRAS.388..247D}. Over the years, numerous signal-separation algorithms have been proposed to address this issue. These techniques are generally classified as either blind or nonblind, depending on whether prior knowledge of the signal, foreground, or instrumental noise is required in the separation process. A comprehensive review of current foreground mitigation methods can be found in~\cite{Liu:2019awk}. Representative blind (or semiblind) approaches include principal component analysis (PCA)~\citep{2011PhRvD..83j3006L}, singular value decomposition (SVD)~\citep{Switzer:2015ria,Zuo:2022wra}, independent component analysis (ICA)~\citep{2012MNRAS.423.2518C,Cunnington21}, and generalized morphological component analysis (GMCA)~\citep{2013MNRAS.429..165C}, all of which exploit the spectral smoothness of foregrounds to isolate the rapidly varying cosmological component. In contrast, nonblind approaches such as the generalized needlet internal linear combination (GNILC)~\citep{2016MNRAS.456.2749O,Marins22}, Gaussian process regression (GPR)~\citep{2012MNRAS.423.2518C,Kern:2020kky}, and the Karhunen-Loève transform~\citep{Shaw:2013wza} incorporate additional statistical priors or external templates, achieving improved separation at the cost of model dependence. In addition, the Bayesian framework \citep{Ghosh:2015fxa, 2016ApJS..222....3Z, Sims:2019iyz, 2023MNRAS.520.4443B} enables joint inference of the 21-cm signal and foregrounds, incorporating prior knowledge and naturally accounting for correlations and uncertainties. This probabilistic approach allows robust separation of the faint cosmological signal from dominant foregrounds while consistently propagating errors.

Furthermore, the telescope’s primary beam introduces additional complexity: its intrinsic chromaticity smooths the sky differently across frequencies, imprinting artificial spectral structures that mimic genuine fluctuations (e.g., \cite{2024ApJS..274...44D}). This beam-frequency coupling reduces the apparent smoothness of the foregrounds, couples angular and spectral modes, and increases the number of PCA modes that must be removed, thereby elevating the risk of losing part of the cosmological signal. Both over- and undersubtraction can bias power-spectrum measurements~\cite{2020MNRAS.495.1788A,2020MNRAS.499.4613S}, and consequently, most existing 21-cm analyses still depend on cross-correlations with external galaxy surveys for robust detection. These challenges highlight the need for improved mitigation strategies that integrate established linear filters such as PCA with advanced reconstruction techniques, enabling more reliable and bias-resistant recovery of the underlying HI signal.

Recent efforts have explored machine-learning approaches to improve signal recovery beyond traditional linear foreground removal (e.g., \cite{2019MNRAS.485.2628L,  2021JCAP...04..081M, 2021MNRAS.504.4716G, 2022ApJ...934...83N, 2024PhRvD.109f3509S, 2024MNRAS.529.3684K, 2024MNRAS.528.5212B, 2025MNRAS.541..234B, 2025MLS&T...6a5039S, 2025arXiv250813265C}). The Deep21 framework \cite{2021JCAP...04..081M} demonstrated that a three-dimensional U-shaped convolutional neural network (UNet) can recover the HI signal with significantly less bias than PCA, leading to improved power-spectrum reconstruction across a wide range of scales and reducing the level of signal loss typical of linear cleaning. Building on this result, Ref.~\cite{2022ApJ...934...83N} investigated the impact of frequency-dependent primary beams and showed that applying a UNet after PCA cleaning can partially restore modes suppressed by aggressive mode subtraction and effectively mitigate beam-induced spectral distortions. More recently, frequency-differencing (FD) combined with deep learning has been proposed as a complementary strategy, where differencing adjacent frequency channels reduces the dynamic range of the input data and enables the network to focus on learning the residual foreground structures \cite{2024PhRvD.109f3509S}. These studies together demonstrate that deep learning can recover cosmological information that would otherwise be lost and that it is emerging as a powerful complement to traditional cleaning techniques.

Despite these advances, key limitations remain that must be addressed before unbiased cosmological measurements become feasible. In particular, it is still unclear how FD-based approaches perform in the presence of realistic, frequency-dependent beams and how they compare systematically with established PCA-based cleaning under identical conditions. Moreover, while PCA robustly removes the bulk of smooth foreground modes, it risks excessive signal loss, whereas deep-learning approaches can recover complex structures but may remain sensitive to residual large-scale contamination. The objective of this work is therefore to close these gaps by systematically testing the FD-based deep-learning method with realistic beam effects, performing a direct performance comparison with PCA cleaning, and exploring a combined approach that leverages the advantages of both techniques. This study aims to clarify the regimes where each method is optimal and to provide a foundation for reliable HI signal recovery, which is essential for precision measurements of BAO, RSD, and HI bias with upcoming low-redshift intensity mapping surveys.

This paper is organized as follows. Sec.~\ref{sec:dataset} introduces the simulation data. In Sec.~\ref{sec:method}, we describe a new foreground removal technique, which combines a hybrid frequency-differencing scheme with a PCA-based deep-learning approach. The performance of the method and the corresponding results are presented in Sec.~\ref{sec:results}. Finally, the main findings are summarized in Sec.~\ref{sec:summary}.

\section{Dataset}
\label{sec:dataset}
In this study, we construct the 21-cm intensity mapping dataset using the same HI and foreground simulations as in our previous work (hereafter Paper I) \cite{2024PhRvD.109f3509S}. We further incorporate a more advanced beam model that closely captures the characteristics of MeerKAT \cite{2016mks..confE..32S}, incorporating frequency-dependent variations and beam-smoothing effects. This refinement improves the realism of our analysis and enables a more robust evaluation of the method’s performance under conditions comparable to those anticipated in forthcoming observations.

\subsection{HI signal and foregrounds}
\label{subsec:hi_fgrd}
For the cosmological 21-cm signal and the foreground components, we maintain the same simulation pipeline as in Paper I, employing the {\tt crime}\footnote{\url{http://intensitymapping.physics.ox.ac.uk/CRIME.html}} code \citep{2014MNRAS.444.3183A}. Below, we briefly summarize these components and refer the reader to Paper I for a detailed description.

The cosmological HI signal is simulated using a log-normal realization of the dark matter density field to model the underlying large-scale structure. The HI overdensity field is then mapped onto spherical shells corresponding to different redshifts, including the effects of RSDs by perturbing the redshifts according to the line-of-sight peculiar velocity. The mean brightness temperature at each redshift is computed as:
\begin{equation}
    \bar{T}_{\mathrm{HI}}(z) = 190.55 \, \mathrm{mK} \, \frac{\Omega_{\mathrm{b}} h (1+z)^2 x_{\mathrm{HI}}(z)}{\sqrt{\Omega_{\mathrm{m}}(1+z)^3 + \Omega_{\Lambda}}},
\end{equation}
where $\Omega_{\mathrm{b}}$, $\Omega_{\mathrm{m}}$, and $\Omega_{\Lambda}$ are the cosmological parameters; $h$ is the dimensionless Hubble parameter; and $x_{\mathrm{HI}}(z)$ is the neutral hydrogen fraction. The HI brightness temperature fluctuation is then given by
\begin{equation}
    T_{\mathrm{HI}}(\hat{n},z) = \bar{T}_{\mathrm{HI}}(z) \left[1 + \delta_{\mathrm{HI}}(\hat{n},z)\right],
\end{equation}
where $\delta_{\mathrm{HI}}(\hat{n},z)$ denotes the HI overdensity at angular position $\hat{n}$ and redshift $z$.

For foregrounds, the dominant components include Galactic synchrotron radiation, Galactic and extragalactic free-free emission, and extragalactic point sources. The Galactic synchrotron emission is modeled using the Haslam 408 MHz map \citep{1982A&AS...47....1H} as a template, extrapolated to the relevant frequencies with a direction-dependent spectral index \citep{2013A&A...553A..96D}. Small-scale structures, including synchrotron, free-free emission, and point sources, are incorporated via Gaussian realizations of the angular power spectra following \citet{2005ApJ...625..575S}. These components are added using parametrized models, with parameters (e.g., amplitude, spectral index, and angular power spectrum) calibrated separately for each component based on observational data. It is worth noting that the Haslam map used in the crime contains residual striping and point-source artifacts, which could lead to inadvertent duplication of point sources in the simulations. This concern can be alleviated by the utilization of an improved version of the Haslam map, which was reprocessed by \citep{2015MNRAS.451.4311R}. Polarized foregrounds, including Faraday rotation effects \citep{1986rpa..book.....R}, are also considered, in accordance with Paper I.

\begin{figure*}
    \centering
    \includegraphics[width=0.7\textwidth]{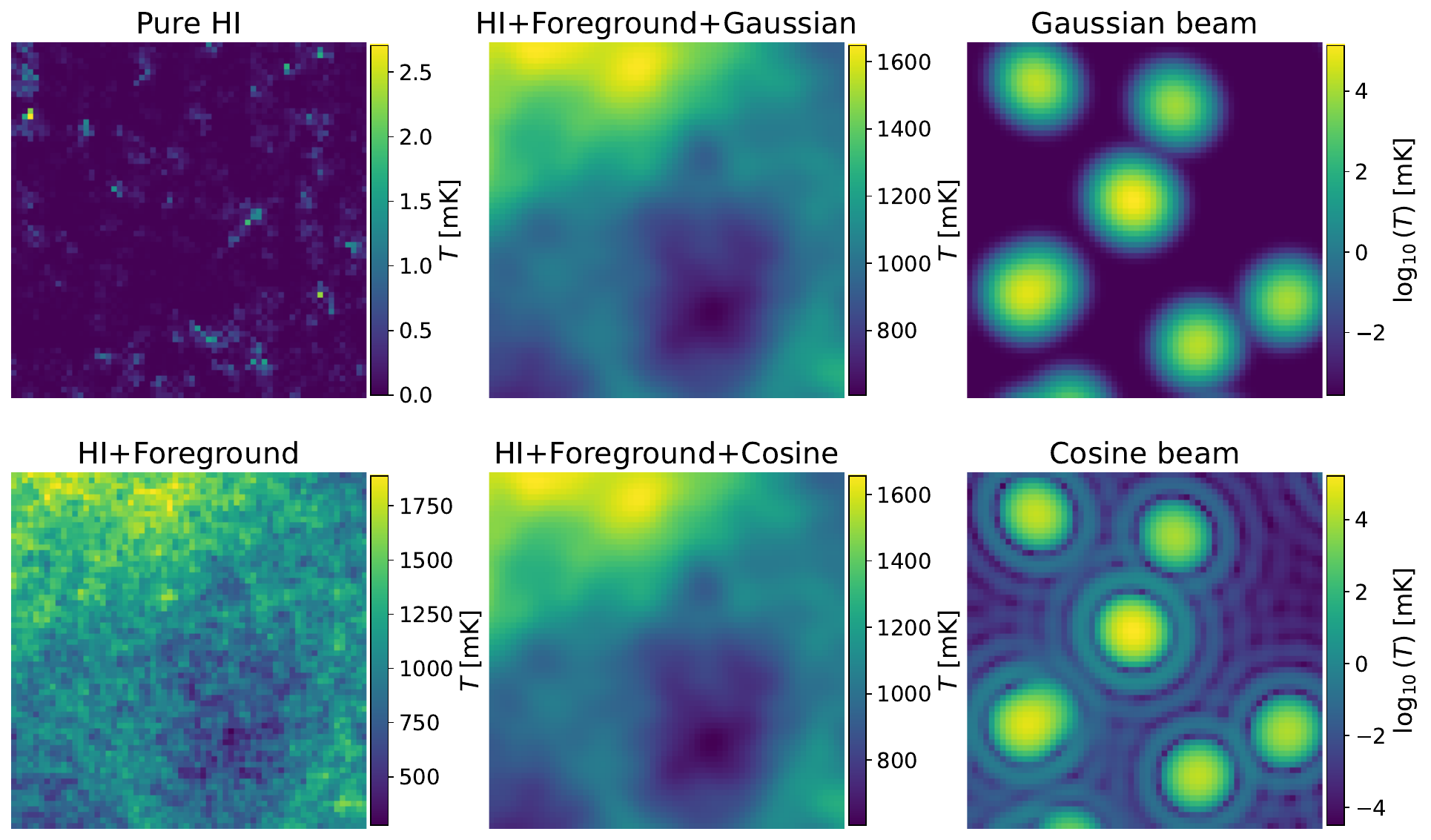}
    \caption{Slices of the simulated data cube at 1100 MHz. Left: pure HI field (top) and HI with foregrounds (bottom). Middle: maps including foreground and convolved with a Gaussian beam (top) and a frequency-dependent cosine beam (bottom). Right: corresponding beam patterns in log scale, illustrated using maps constructed from randomly distributed point sources to highlight the beam structure.}
    \label{fig:beam_2d}
\end{figure*}

In this study, we use a map with $N_{\rm side}=256$ in the {\tt HEALPix} pixelization scheme~\citep{2005ApJ...622..759G}, corresponding to an angular resolution of 13.73 arc min. We consider a frequency range spanning from 1100 to 1164 MHz, which corresponds to a redshift range between 0.29 and 0.22.
The simulated full sky is subsequently divided into 192 localized patches, each covering approximately $214.86~ \mathrm{deg}^2$ sky area. For each patch, we extract a data cube with dimensions $64^3$, representing two spatial dimensions and one frequency axis. The left panels of Fig.~\ref{fig:beam_2d} show slices of the simulated HI signal (top) and foreground-contaminated maps (bottom) at 1100 MHz. Each cube corresponds to a comoving volume approximately equivalent to a cube with a side length of 
$187 \mpch$. The cosmological parameters are set as follows: $\Omega_m=0.3$, $\Omega_\Lambda=0.7$, $\Omega_b=0.049$, $h=0.67$, and $\sigma_8=0.8$.

\begin{figure*}
    \centering
    \includegraphics[width=0.8\linewidth]{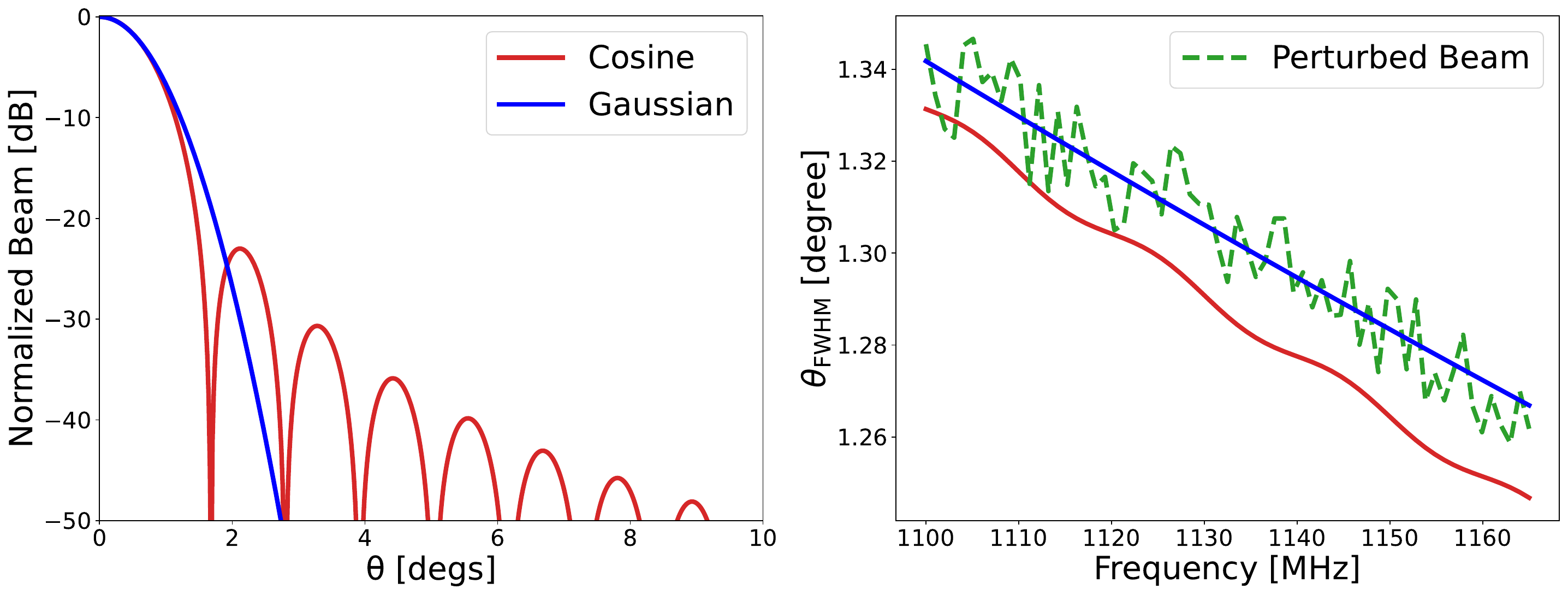}
   \caption{Left: normalized angular beam profiles of the Gaussian (blue) and cosine (red) beam models as a function of angular separation from the beam center at 1100 MHz. Right: frequency dependence of the FWHM for the Gaussian (blue) and cosine (red) beam models. The green dashed curve represents a perturbed Gaussian beam with small frequency-dependent perturbations, used to model beam uncertainties in the mismatch test (test A in Sec.~\ref{sec:mismatch}).}
    \label{fig:beam_profile}
\end{figure*}

\subsection{Instrumental beam effects}
To simulate the realistic observational effect, we incorporate two distinct primary beam models into our analysis: a Gaussian beam model and a more sophisticated cosine beam model. These models are implemented to evaluate the impact of instrumental effects on the observed sky maps and to test the robustness of foreground removal methods under increasingly realistic conditions.

The Gaussian beam model serves as an idealized, simplified approximation of the primary beam response. It is modeled as a symmetric, frequency-dependent function,
\begin{equation}
B_G(\nu, \theta) = \exp \left[ -4 \ln 2 \left( \frac{\theta}{\theta_{\mathrm{FWHM}}(\nu)} \right)^2 \right],
\end{equation}
where $\theta$ is the angular distance from the beam center, and $\theta_{\mathrm{FWHM}}(\nu)$ is the full width at half maximum (FWHM) of the beam at frequency $\nu$. The FWHM is calculated as,
\begin{equation}
\theta_{\mathrm{FWHM}}(\nu) = 1.16 \frac{\lambda(\nu)}{D},
\end{equation}
where $\lambda(\nu) = c/\nu$ is the wavelength at frequency $\nu$, $c$ is the speed of light, and $D$ is the dish diameter, which we set to $D = 13.5\, \mathrm{m}$ to match MeerKAT specifications.

To achieve a more realistic representation of the primary beam, we adopt a cosine beam model that captures both the mainlobe and the frequency-dependent ripple structures observed in holographic measurements of the MeerKAT beam pattern \cite{2021MNRAS.506.5075M}. The beam pattern is expressed as:
\begin{equation}
B_C(\nu, \theta) = \left[ \frac{\cos \left(1.189\pi \theta / \Delta \theta_C(\nu)\right)}{1 - 4 \left(1.189 \theta / \Delta \theta_C(\nu)\right)^2} \right]^2,
\end{equation}
where $\Delta \theta_C(\nu)$ represents the frequency-dependent effective beam width of the cosine model. The effective beam width is modeled as:
\begin{equation}
\Delta \theta_C(\nu) = \frac{\lambda(\nu)}{D} \left[ \sum_{d=0}^{8} a_d \, \hat{\nu}^d + A \sin \left( \frac{2\pi \hat{\nu}}{T} \right) \right],
\end{equation}
where $\hat{\nu} = \nu / \mathrm{MHz}$ is the normalized frequency, $A=0.1$ and $T=20$ define the amplitude and period of the sinusoidal ripple component, and the polynomial coefficients are set to: $a_d = \{3.402 \times 10^{-21}, -3.025 \times 10^{-17}, -1.17 \times 10^{-11}, -2.572 \times 10^{-10}, 3.511 \times 10^{-7}, -3.05 \times 10^{-4}, 0.1645, -50.37, 6704.28\}$, which are  empirically derived from MeerKAT beam fitting data.

The middle panels of Fig.~\ref{fig:beam_2d} show slices of the beam-convolved maps, using a Gaussian beam (top) and a frequency-dependent cosine beam (bottom). To illustrate the differences between the Gaussian and cosine beam models, the right panels of Fig.~\ref{fig:beam_2d} show the corresponding beam patterns in log scale, constructed using maps of randomly distributed point sources. After convolution, the Gaussian beam produces smooth and compact source profiles, whereas the cosine beam introduces pronounced ringlike patterns and extended sidelobe structures that can mimic diffuse foreground emission. This example highlights the importance of realistic beam modeling for reliable foreground removal. The left panel of Fig.~\ref{fig:beam_profile} presents the normalized angular beam patterns at 1100 MHz. The cosine beam exhibits pronounced sidelobes that extend to larger angular scales, whereas the Gaussian beam decreases smoothly without significant sidelobes. These sidelobes can inject additional power on large angular scales, thereby complicating the separation of foregrounds from the cosmological 21-cm signal. The right panel of Fig.~\ref{fig:beam_profile} shows the frequency dependence of the FWHM for both models. The Gaussian beam maintains a systematically larger FWHM due to its simple, symmetric profile that neglects detailed beam structures. In contrast, the cosine beam captures realistic mainlobe narrowing and frequency-dependent distortions, offering a more accurate representation of instrumental chromaticity that is critical for foreground mitigation.

In addition to the Gaussian and cosine beam models, we introduce a perturbed Gaussian beam to mimic realistic beam uncertainties. Specifically, the FWHM of the perturbed beam is modeled as
\begin{equation}\label{eq:pertbeam}
\theta_{\rm FWHM}^{\rm pert}(\nu)
= \theta_{\rm FWHM}^{\rm Gauss}(\nu)\,[1 + \delta(\nu)], 
\end{equation}
where $\delta(\nu)$ represents a small, frequency-dependent random perturbation drawn independently at each frequency. We adopt a fluctuation level of approximately 1\% to mimic residual errors in beam calibration and modeling, which are typically at the percent level in single-dish HI intensity mapping experiments (e.g., \cite{2016ApJ...833..289L}). As illustrated by the green dashed line in Fig.~\ref{fig:beam_profile}, this construction preserves the overall Gaussian beam profile while introducing realistic fluctuations in its chromatic evolution. This perturbed beam is used in the beam-mismatch test (test A) described in Sec.~\ref{sec:mismatch}.

The inclusion of these beam models in our simulations allows for a thorough assessment of foreground mitigation techniques, ensuring that our analysis accounts for instrumental effects ranging from simplified to more complex cases, in alignment with actual observations from the MeerKAT telescope.
\begin{figure*}
    \centering
    \includegraphics[width=0.9\linewidth]{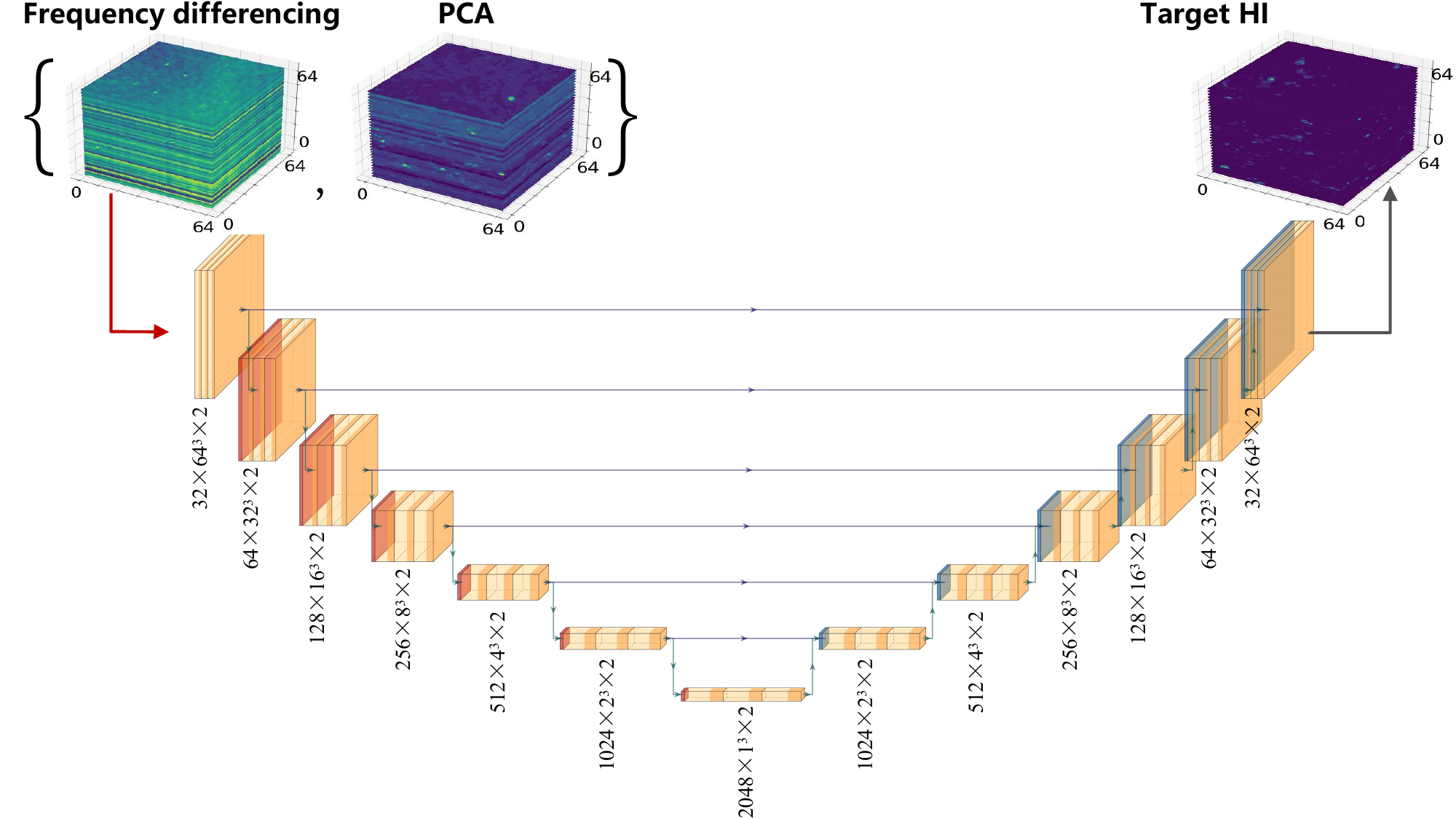}
    \caption{Visualization of the UNet architecture. The input is a cube of size $64^3$ with two channels derived from the frequency-difference and PCA-cleaned maps. The architecture consists of 13 layers, with 6 encoder layers, 6 decoder layers, and 1 bottleneck layer. Each layer includes three convolution operations with batch normalization and activation functions (except for the output layer).  The annotations below each layer (e.g., $32 \times 64^2 \times 2$) denote the number of feature maps, the spatial dimension, and the two-channel processing. Down, right, and up arrows represent max pooling, skip connections, and transpose convolutions, respectively. This visualization was created using the \texttt{PlotNeuralNet} library. }
    \label{fig:Unet}
\end{figure*}
\begin{figure}
    \centering
    \includegraphics[width=1.03\linewidth]{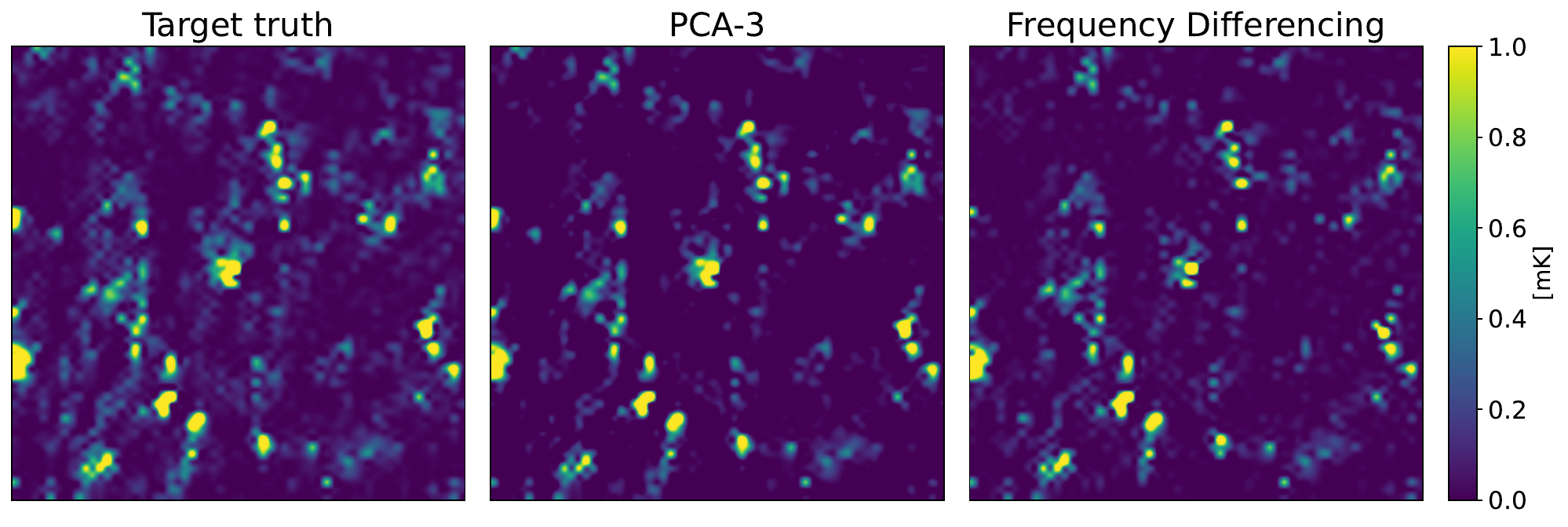}
    \caption{Comparison of preprocessing methods applied to pure HI signal fields. From left to right, the panels show the target truth, the PCA-3 reconstruction, and the frequency-differencing result, respectively.}
    \label{fig:slice_preprocess}
\end{figure}

\begin{figure}
    \centering
    \includegraphics[width=1.01\linewidth]{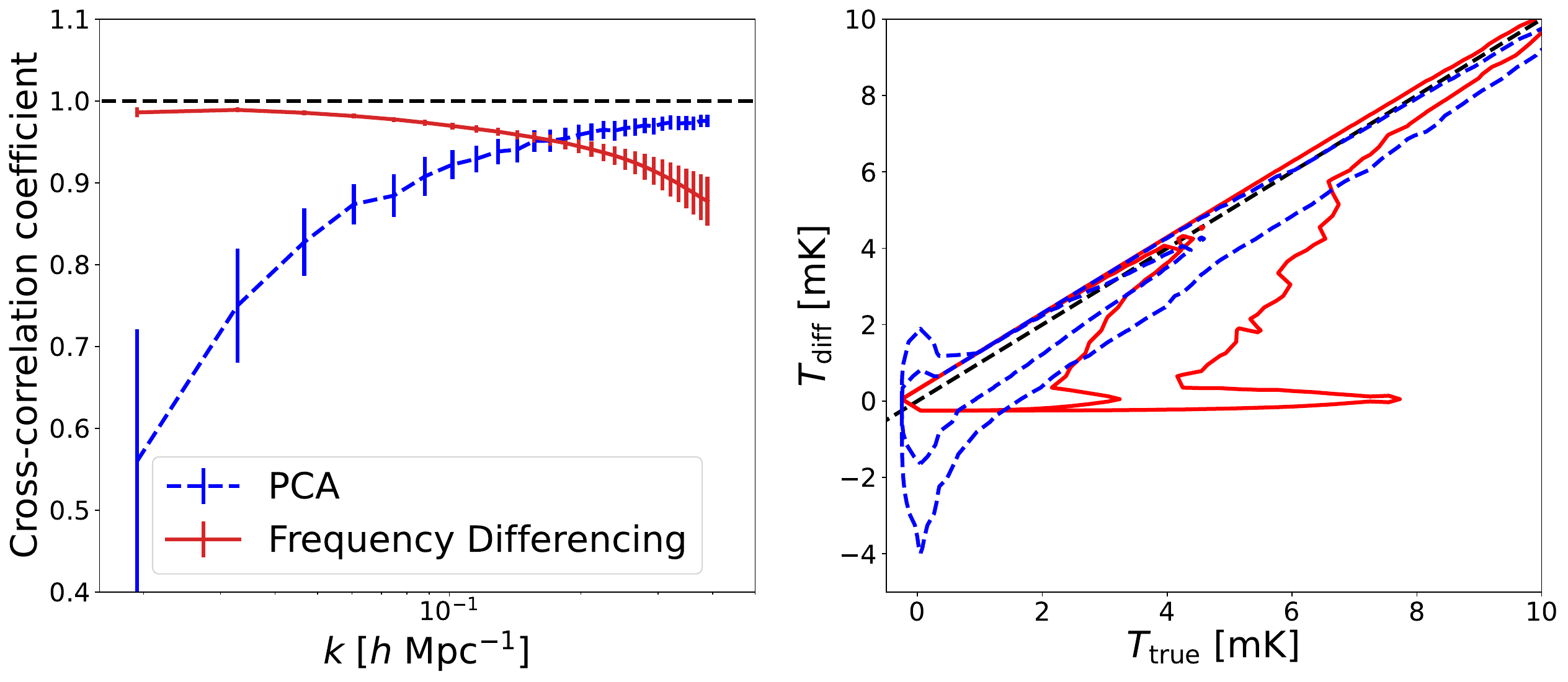}
    \caption{Comparison of preprocessing methods applied to the pure HI fields. Left: cross-correlation coefficient $R_{\text{cross}}(k)$ between the reconstructed and true HI signals. Right: pixel-to-pixel correlation of the preprocessed temperature $T_{\mathrm{diff}}$ with the true temperature $T_{\mathrm{true}}$, where the contours enclose 65\% and 95\% of the grid cells.}
    \label{fig:cross_preprocess}
\end{figure}

\section{FOREGROUND REMOVAL METHOD}\label{sec:method}

In this section, we describe the foreground removal strategy employed in this study, which is based on deep-learning techniques. Specifically, we use a UNet architecture and investigate several preprocessing methods designed to improve its performance in the presence of foreground contamination and instrumental beam effects.

\subsection{UNet architecture}
We adopt a UNet architecture to reconstruct the cosmological 21-cm signal from foreground-contaminated and beam-convolved intensity maps. In this study, we use the same UNet structure introduced in Paper I, where a detailed description of the model configuration is provided. For completeness, we briefly summarize the key features here, with emphasis on the modifications for a two-channel input, although the design can be straightforwardly adapted to a single-channel case.

Fig.~\ref{fig:Unet} presents a visualization of the UNet model architecture. The network input is a cube of size
$64^3$ with two channels obtained from data preprocessing, which will be introduced in the next subsection. The output is a single-channel HI cube of the same size. The encoder comprises six downsampling stages that progressively increase the number of feature maps from 32 to 2048, enabling the network to distinguish spectrally smooth foreground components from the rapidly varying HI signal. The decoder symmetrically upsamples these representations to recover the target field, thereby recovering the cosmological signal while suppressing residual contamination. Each block applies a pair of successive 3D convolutional layers, and the annotations in the figure (e.g., $32 \times 64^2 \times 2$) denote the number of feature maps, the spatial dimension, and the two-channel processing. Skip connections between corresponding encoder and decoder layers preserve spatial information and enable the network to recover fine-scale HI features that might otherwise be lost during downsampling. Through this structure, the UNet implicitly learns to disentangle the spectrally smooth foregrounds from the cosmological signal during training.

The UNet is trained in a supervised manner, with the loss function defined as the mean squared error between the network output and the ground-truth HI-only signal cube. The network processes 102 training samples each epoch, followed by 20 validation samples and 70 testing samples. Finally, we save our best model based on the minimum validation-set loss function.

\subsection{Preprocessing strategies}
The UNet is expected to learn the separation of foregrounds from the 21-cm signal by leveraging differences in their spectral smoothness and spatial coherence. However, applying the UNet directly to raw data with full foreground complexity remains challenging, particularly under large dynamic range conditions. As demonstrated in Paper I and other studies (e.g., \cite{2021JCAP...04..081M, 2022ApJ...934...83N}), 
the UNet-based method critically depends on preprocessing strategies that reduce the foreground amplitude or decorrelate its dominant spectral modes. To address this challenge, we explore three preprocessing techniques prior to UNet training, as described below.

\subsubsection{Principal component analysis}

PCA is used to exploit the spectral smoothness of foregrounds compared to the rapidly fluctuating cosmological signal. In our case, PCA is applied to each of the 192 simulated sky patches, with each $64^3$ data cube reshaped into a collection of one-dimensional spectra along the frequency axis. A frequency-frequency covariance matrix is computed across all spectra in each patch, serving as the basis to identify the dominant spectral modes. The resulting principal components, associated with the largest eigenvalues, are interpreted as the foreground structure and subtracted to produce a cleaned residual cube.

We subtract a fixed number of principal components (typically 3) from each frequency spectrum to produce a residual cube that suppresses most foreground power while retaining a substantial fraction of the 21-cm signal. Following {\tt Deep21}, the PCA-cleaned map is then used as input to the UNet for both training and inference. While PCA does not require any foreground modeling, it can inadvertently remove cosmological signal on large scales, necessitating careful evaluation of signal loss.

\subsubsection{Frequency differencing}

Frequency differencing is a model-independent technique that exploits the spectral smoothness of foregrounds relative to the cosmological 21-cm signal. By taking differences between adjacent frequency channels along each line of sight, the spectrally coherent foreground components, which vary slowly with frequency, are effectively suppressed, while the rapidly fluctuating 21-cm signal is largely preserved.

In our implementation, the differenced cube is generated by computing the finite difference along the frequency axis,
\begin{equation}
\delta T(\nu_i) = T(\nu_{i+1}) - T(\nu_i),
\end{equation}
where \(T(\nu_i)\) denotes the brightness temperature at frequency \(\nu_i\). This operation results in a residual map that significantly reduces the amplitude of foregrounds without requiring any explicit modeling or eigendecomposition.

In this study, we adopt a fixed frequency resolution of 1 MHz. As demonstrated in Paper I, the choice of differencing bandwidth plays a critical role in balancing foreground suppression and signal preservation. Narrower bandwidths tend to enhance the subtraction of spectrally smooth foregrounds but can amplify noise and reduce sensitivity to the HI signal. Conversely, broader bandwidths preserve more signal but may allow residual foreground contamination to remain. Optimal performance is therefore achieved by tuning the differencing scale according to the specific characteristics of the signal and noise in the data. In our case, the 1 MHz spacing provides a reasonable trade-off given the resolution and noise properties of the simulated data. 

Moreover, it is important to consider that the size of the beam ($\theta_\mathrm{FWHM}$) varies with frequency. To ensure consistency in the beam size level, we have smoothed each adjacent-band map to match their relatively lower angular resolution. This was done by convolving the higher frequency map with a Gaussian beam characterized by a beam size of $\Delta\theta_\mathrm{FWHM}$, denoted as,
\begin{equation}\label{eq:smooth}
    \Delta\theta_\mathrm{FWHM} = \sqrt{\theta_\mathrm{FWHM,target}^2-\theta_\mathrm{FWHM,original}^2}~.
\end{equation}
Here $\theta_\mathrm{FWHM, target}$ and $\theta_\mathrm{FWHM, original}$ correspond to the beam size of low and high frequency, respectively. It reveals that the frequency-difference map yields a similar outcome, with a significant reduction in the range of amplitude.

In this work, we treat the differenced cube as an alternative input to the UNet and evaluate its performance relative to PCA-based preprocessing in terms of signal recovery and residual foreground suppression.

\begin{figure*}
    \centering
    \includegraphics[width=1.01\linewidth]{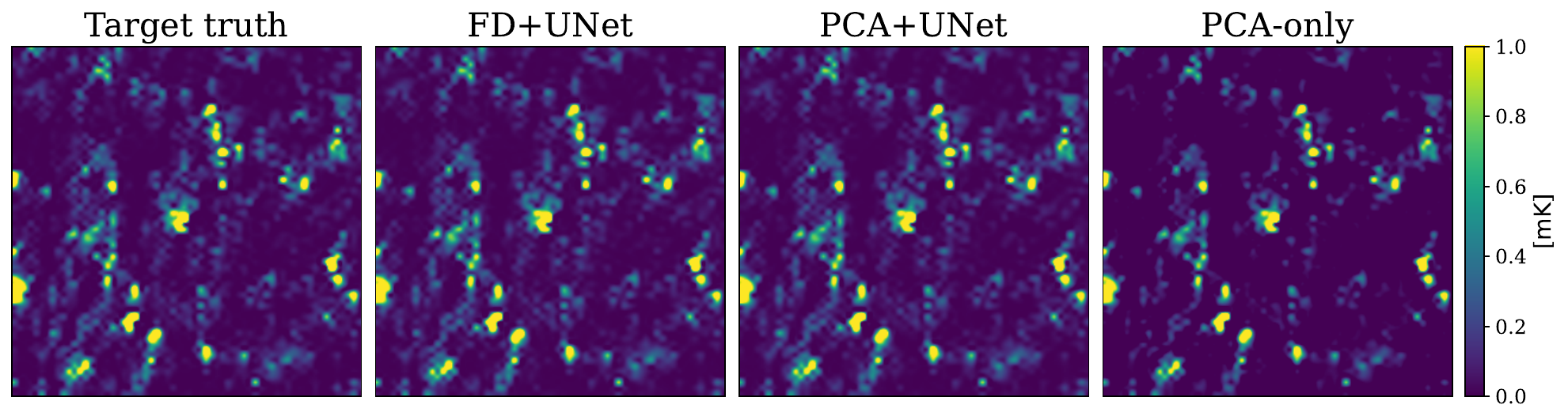}
    \caption{Comparison of reconstructed HI maps without beam effects. From left to right, the panels show the true HI signal (target truth), the FD+UNet reconstruction, the PCA+UNet reconstruction, and the PCA-only reconstruction.}
    \label{fig:slice_nobeam}
\end{figure*}
\begin{figure*}
    \centering
    \includegraphics[width=0.93\linewidth]{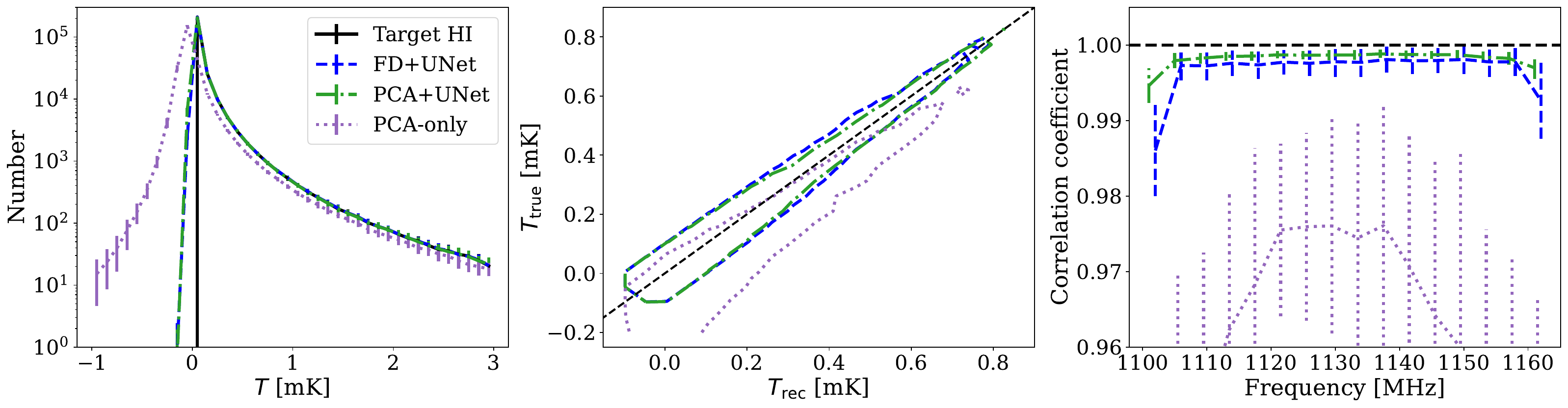}
    \caption{Diagnostics of reconstruction performance without beam effects. Left: temperature distribution of the reconstructed and true HI fields. Middle: pixel-to-pixel correlation between reconstructed and true fields, with contours enclosing about 95\% of the grid cells. Right: correlation coefficients as a function of frequency. The blue dashed, green dash-dotted, and purple dotted lines correspond to the FD+UNet, PCA+UNet, and PCA-only cases, respectively. Error bars indicate the $1\sigma$ scatter across the testing samples.}
    \label{fig:pixel_nobeam}
\end{figure*}

\subsubsection{Hybrid method: Dual-channel PCA and frequency differencing}

To leverage the complementary strengths of PCA and frequency differencing, we construct a hybrid preprocessing approach in which the outputs of both methods are used in parallel. We preprocess the input data cube independently using PCA and frequency differencing, yielding two separate residual maps. These are then combined to form a two-channel input for the UNet.

To motivate the hybrid preprocessing approach, we compare the performance of PCA and frequency differencing by applying each method to pure HI signal maps that do not include any foreground. As shown in Fig.~\ref{fig:slice_preprocess}, the PCA reconstruction (middle panel) preserves most of the bright small-scale structures but also suppresses part of the diffuse background, leading to a noticeable loss of large-scale information. In contrast, the frequency-differencing method (right panel) better retains the diffuse large-scale features but introduces residual striping and artificial patterns, which distort small-scale structures. This complementary behavior highlights the potential of a hybrid preprocessing strategy, which aims to combine the strengths of both methods while mitigating their respective limitations.

Furthermore, Fig.~\ref{fig:cross_preprocess} shows the cross-correlation coefficient between the recovered and true HI fields (left panel) and the pixel-to-pixel temperature correlation (right panel). The cross-correlation results show that frequency differencing (red) achieves nearly perfect correlation on large scales ($k \lesssim 0.1 ~h~\mathrm{Mpc}^{-1}$), while PCA (blue) significantly underestimates the signal due to the removal of large-scale modes. However, at smaller scales, PCA recovers the correlation more accurately, whereas frequency differencing gradually loses performance as residual distortions dominate. The pixel-to-pixel correlation confirms that the PCA reconstruction (blue) scatters more broadly around the one-to-one line, especially at low temperatures, which indicates a loss of diffuse components. By contrast, frequency differencing (red) follows the one-to-one relation more closely for diffuse emission but introduces systematic deviations and distortions for bright pixels. These results highlight the complementary nature of the two methods, with PCA being more reliable at preserving small-scale structures and frequency differencing more effective at retaining large-scale diffuse features.
 
These complementary behaviors highlight the benefit of the hybrid strategy, in which both PCA-cleaned and frequency-differenced maps are used as two-channel inputs to the UNet. This configuration enables the network to simultaneously exploit the large-scale accuracy of frequency differencing and the improved small-scale fidelity of PCA-based suppression. We assess the effectiveness of this dual-channel hybrid method by comparing its performance against the single-channel PCA-only and frequency-differencing-only configurations in terms of signal recovery and residual foreground contamination.

\section{Results}\label{sec:results}

In this section, we evaluate the performance of the
UNet-based foreground removal method under a set of
increasingly realistic observational conditions. Specifically,
we consider four cases: 
\begin{itemize}
    \item case I: input maps containing only the 21-cm signal and foregrounds;
    \item case II: the same maps as in case I, but
    convolved with a Gaussian instrumental beam;
    \item case III: the same maps as in case I, but convolved with a more realistic, frequency-dependent cosine beam profile;
    \item case IV: beam-mismatch tests, in which the beam model used in the testing data differs from that adopted in training.
\end{itemize}

For case I, we compare two preprocessing strategies,
namely PCA and frequency differencing (FD). For cases
II--IV, we additionally include their combination
(hybrid) and compare the performance of PCA, FD, and
hybrid in recovering the underlying HI signal. Results from
the PCA-only method are included as a baseline reference.
For all PCA-related approaches, three principal components
are removed to suppress the dominant smooth spectral
modes.

Moreover, we note that, for the first three cases, an independent
network is trained and tested under the same beam
assumption, i.e., the beam model used to generate the
training set is identical to that used in the testing data.
For the last case, however, this matched condition is intentionally
relaxed, such that the beam model used in the testing data
differs from that adopted in training.

\begin{figure*}
    \centering
    \includegraphics[width=0.8\linewidth]{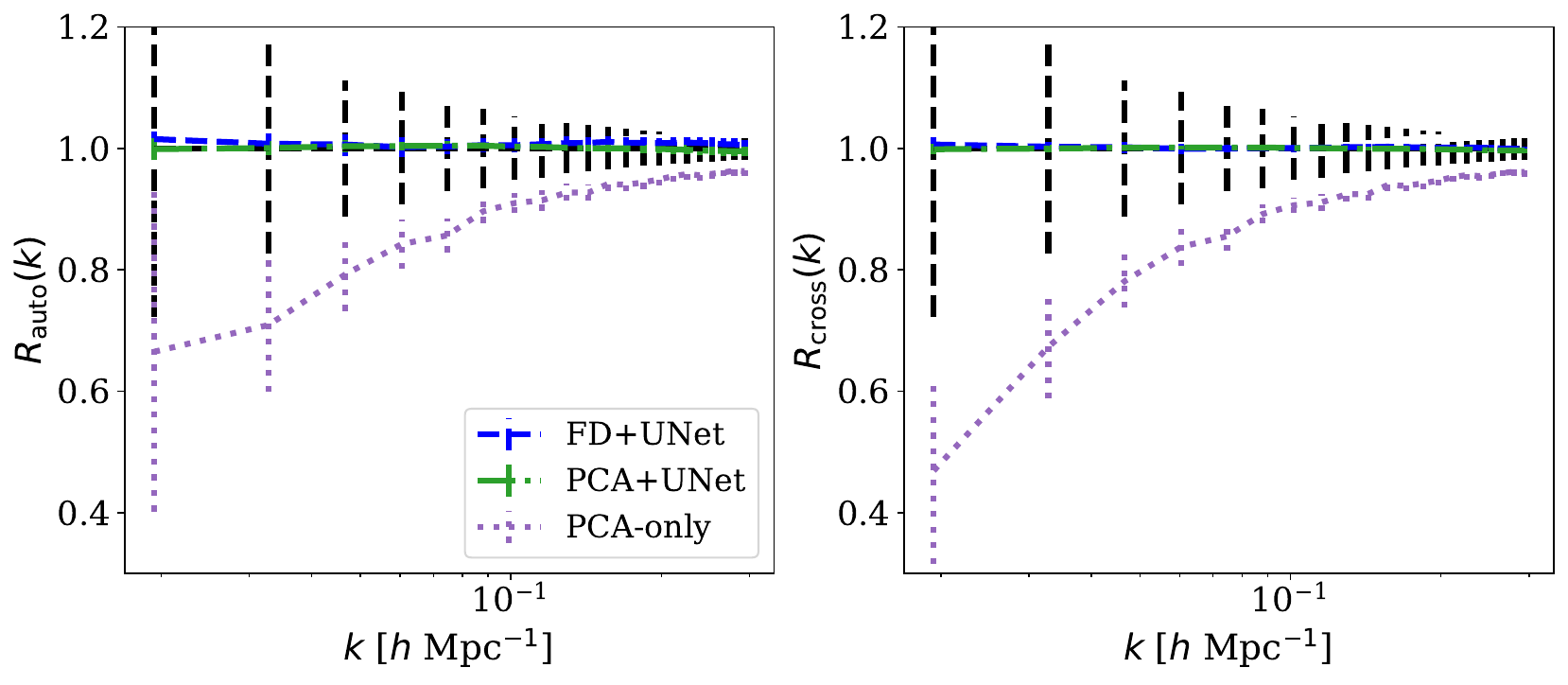}
    \caption{Auto- (left) and cross-power (right) spectrum ratios without beam convolution. The results of FD+UNet (blue dashed), PCA+UNet (green dash-dotted), and PCA-only reconstruction (purple dotted) are shown for comparison. The black dashed error bars represent the theoretical uncertainties computed using Eq.~(\ref{eq:pker}).}
    \label{fig:pk1d_nobeam}
\end{figure*}
\begin{figure*}
    \centering
    \includegraphics[width=1\linewidth]{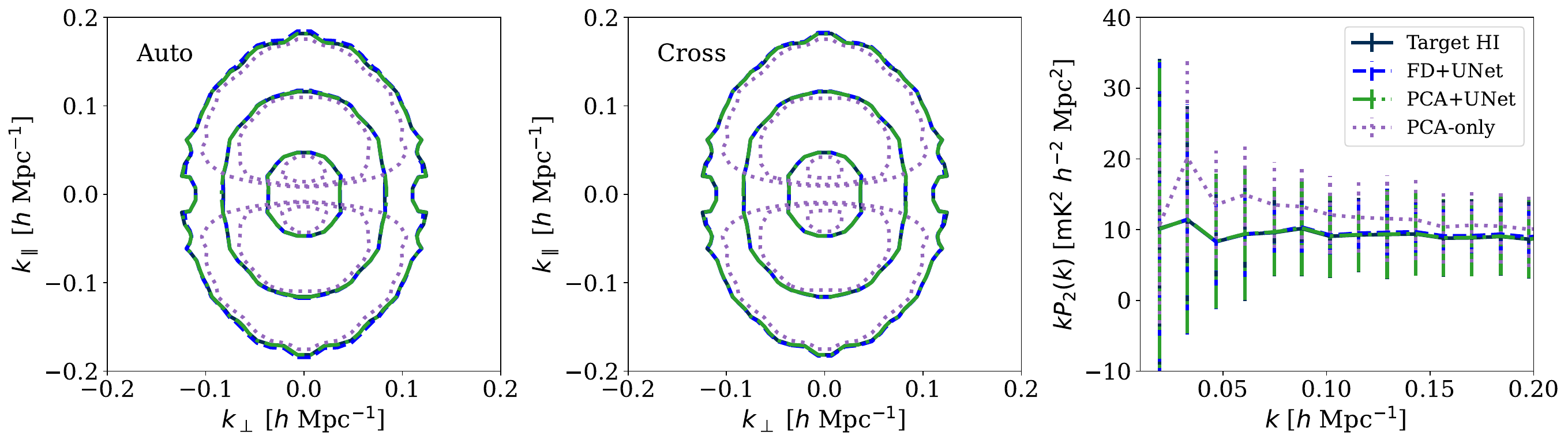}
    \caption{Comparison of 2D power spectra without beam effects. Left: autopower spectra of the reconstructed HI fields. Middle: cross-power spectra between the reconstructed and true HI fields. Right: quadrupole power spectrum of the autocorrelation.  Different lines correspond to the target HI field (black solid), FD+UNet (red dashed), PCA+UNet (blue dash-dotted), and PCA-only (green dotted), respectively. All results are averaged over the test set, with error bars indicating the $1\sigma$ scatter among samples.} 
    \label{fig:pk2d_nobeam}
\end{figure*}

\subsection{HI + foreground: Ideal instrument case}

We begin our analysis with the first case. Fig.~\ref{fig:slice_nobeam} presents two-dimensional (2D) slices of the target HI field alongside the reconstructed fields obtained using the FD+UNet, PCA+UNet, and PCA-only methods, as labeled above each panel. Both UNet-based approaches yield reconstructions that are visually consistent with the ground truth, while the PCA-only result shows a noticeable suppression of temperature distribution due to the aggressive removal of spectral modes.

Fig.~\ref{fig:pixel_nobeam} presents a quantitative comparison across several statistical diagnostics. From left to right, the panels show the temperature distributions, the pixel-to-pixel correlation between reconstructed and true fields, and the correlation coefficient as a function of frequency. Both FD+UNet and PCA+UNet reproduce the HI histogram with high fidelity over the entire dynamic range, whereas the PCA-only reconstruction is systematically biased toward lower amplitudes. The pixel-level comparison further indicates that the two UNet-based models preserve an almost perfectly linear relation relative to the target temperatures, while the PCA-only case shows a biased correlation with systematically reduced amplitudes. The correlation coefficients remain above 0.99 across the full band for both UNet-based approaches, confirming their stable and accurate recovery of the HI signal, in contrast to the substantially degraded correlations produced by the PCA-only method. A minor decline at the band edges appears in the FD+UNet case, which arises from boundary effects introduced by the finite-differencing operation. Because the first and last channels lack symmetric neighboring information, the differencing step cannot be applied consistently at those locations, resulting in reduced accuracy near the frequency boundaries.

Fig.~\ref{fig:pk1d_nobeam} compares the ratios of the autopower spectrum (left) and cross-power spectrum (right) between the recovered and true fields. Here, the auto- and cross-power spectra probe complementary aspects of the reconstruction quality. The autopower spectrum is sensitive to the recovery of the fluctuation amplitude, whereas the cross-power spectrum more directly measures the phase consistency between the reconstructed field and the true HI signal. For this reason, we primarily use the cross-power spectrum to assess the fidelity of the reconstruction, while the autopower spectrum serves as a complementary diagnostic of amplitude-related biases. For the recovered $P(k)$, the error bars represent the $1\sigma$ variance estimated from the test samples. The black error bars on the reference line are derived from the theoretical variance associated with the finite number of Fourier modes in a limited volume \cite{2021PhRvD.104d3528S}, given by
\begin{equation}\label{eq:pker}
\Delta P(k) = \frac{P(k)}{\sqrt{N_k}} ,
\end{equation}
where $N_k$ is the number of independent $k$-modes in each bin. Both FD+UNet and PCA+UNet accurately recover the power spectra across the full $k$-range. By contrast, PCA alone performs poorly at all scales, especially at low $k$, where it suffers significant signal loss.

\begin{figure*}
    \centering
    \includegraphics[width=1.01\linewidth]{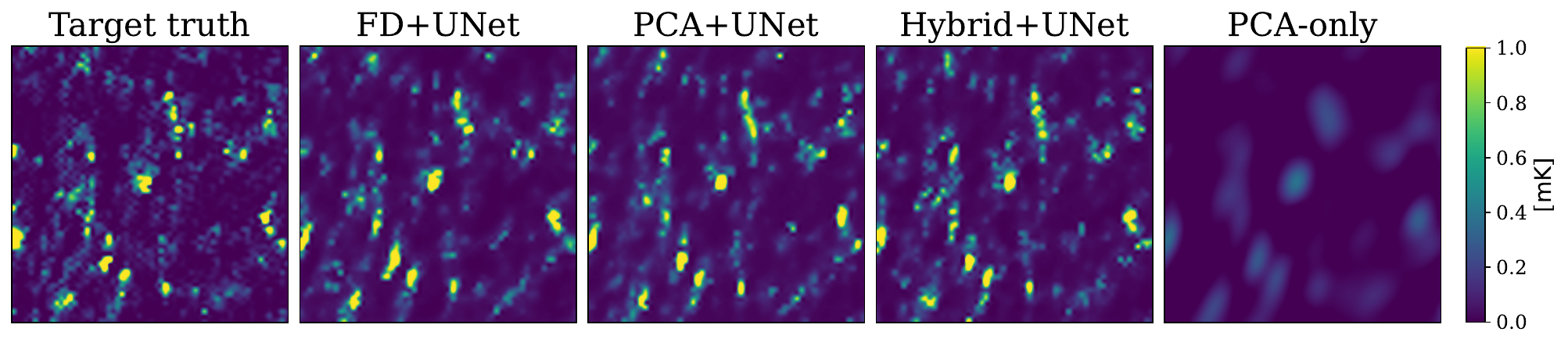}
    \caption{Same as Fig.~\ref{fig:slice_nobeam}, but with Gaussian beam convolution applied. The results from the hybrid method are also included for comparison.}
    \label{fig:slice_gauss}
\end{figure*}
\begin{figure*}
    \centering
    \includegraphics[width=0.97\linewidth]{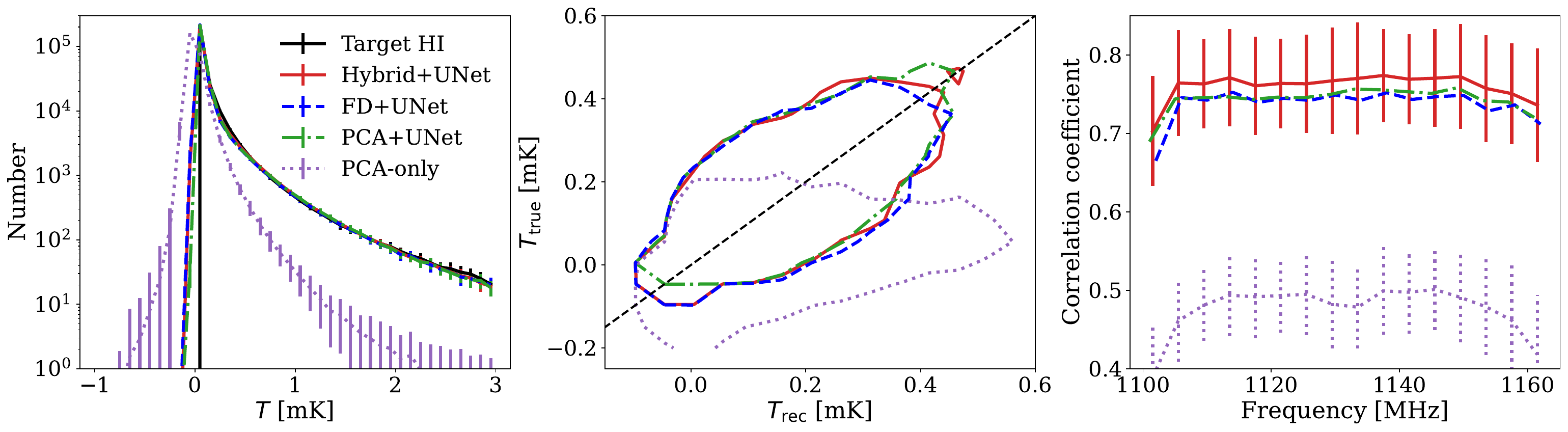}
    \caption{Same as Fig.~\ref{fig:pixel_nobeam}, but with Gaussian beam convolution applied. The results from the hybrid method are also included for comparison. For clarity, error bars are shown only for the hybrid and PCA-only cases in the right panel.}
    \label{fig:pixel_gauss}
\end{figure*}
\begin{figure*}
    \centering
    \includegraphics[width=0.95\linewidth]{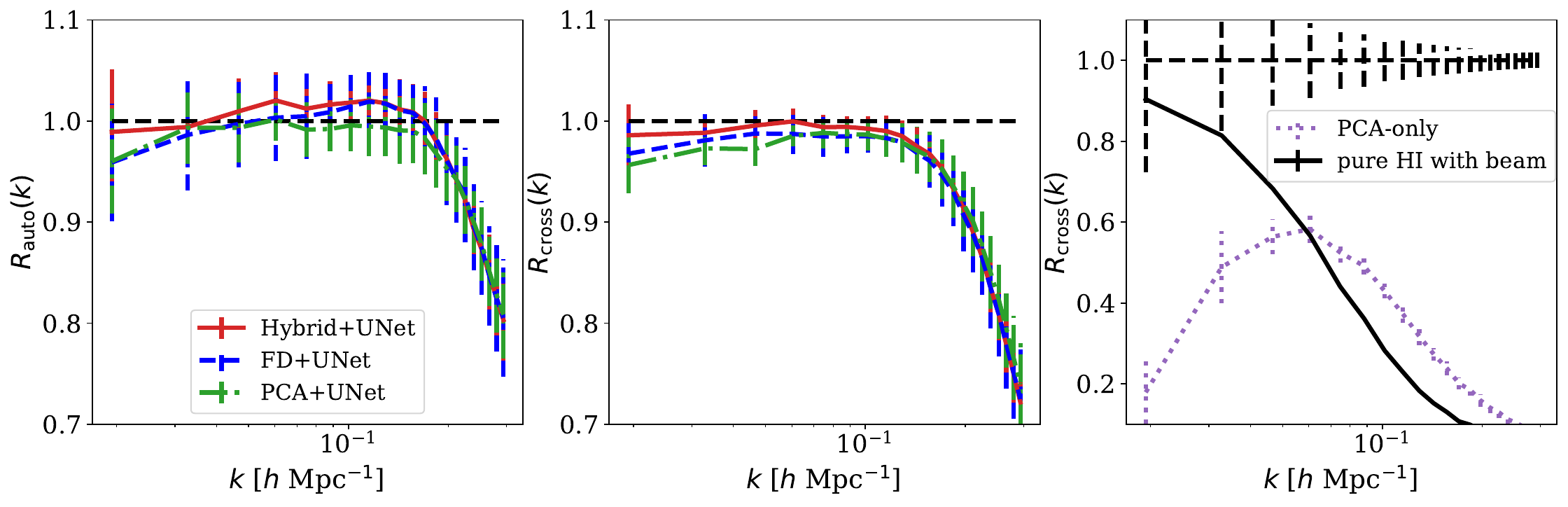}
    \caption{Auto- and cross-power-spectrum ratios under Gaussian beam convolution. The first two panels show the results of the hybrid+UNet (red solid), FD+UNet (blue dashed), and PCA+UNet (green dash-dotted) models, while the third panel presents the pure HI field convolved with the Gaussian beam (black solid) and the PCA-only reconstruction (purple dotted) for reference. The colored error bars indicating the $1\sigma$ variance across the samples, and the black dashed error bars are computed using Eq.~(\ref{eq:pker}).}
    \label{fig:pk1d_gauss}
\end{figure*}

Finally, Fig.~\ref{fig:pk2d_nobeam} shows the 2D power spectrum in the (\(k_\perp\), \(k_\parallel\) ) plane for the auto (left) and cross (middle) spectra. This is used to evaluate how accurately the reconstructed field retains the anisotropic RSD information. Both FD+UNet and PCA+UNet yield power spectra that closely trace the shape and amplitude of the target HI field, while the PCA-only method again shows significant power suppression, especially near the low-\(k_\parallel\) modes where smooth foregrounds dominate.

The Kaiser effect induces a nonzero quadrupole moment $P_2(k)$ to deviate significantly from zero, which serves as a key observable for constraining the growth rate of cosmic structure. The right panel of Fig.~\ref{fig:pk2d_nobeam} compares $P_2(k)$ for the autocorrelation case, which shows a trend similar to that in the cross-correlation. Error bars indicate the $1\sigma$ scatter among the test samples. Both FD+UNet and PCA+UNet successfully recover a nonzero quadrupole signal consistent with the target HI field. Although there is a large uncertainty due to the cosmic variance in our small-volume samples, we can still observe a discrepancy between the PCA reconstructed and the true $P_2(k)$.

In this idealized case without instrumental convolution, both FD+UNet and PCA+UNet deliver high-fidelity reconstructions of the 21-cm signal, significantly outperforming the PCA-only baseline. The UNet is capable of recovering signal components that are partially lost during the preprocessing step by learning spatial and spectral patterns from the training data. Since the
foregrounds in this case remain spectrally smooth and can
already be effectively handled by PCA or FD alone,  the
hybrid method is not expected to provide additional
insight here and is therefore not included. These results validate the effectiveness of applying artificial-intelligence-based methods under simple conditions and establish a foundation for more complex scenarios involving instrumental systematics.

\begin{figure*}
    \centering
    \includegraphics[width=1\linewidth]{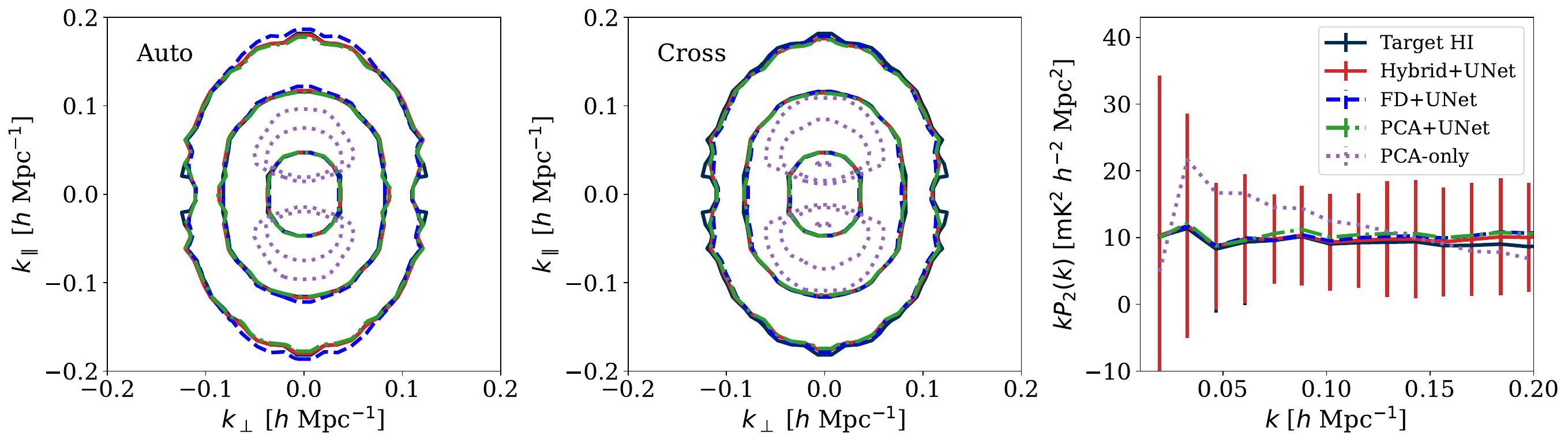}
    \caption{Same as Fig.~\ref{fig:pk2d_nobeam}, but with Gaussian beam convolution applied. The results from the hybrid method are also included for comparison. For clarity, error bars are shown only for the hybrid case in the right panel.}
    \label{fig:pk2d_gauss}
\end{figure*}
\subsection{HI + foreground with Gaussian beam convolution}

We now evaluate the performance of the methods in the presence of instrumental smoothing by convolving the HI+foreground maps with a symmetric Gaussian beam. Fig.~\ref{fig:slice_gauss} shows 2D slices of the target truth and the reconstructed HI fields from  FD+UNet, PCA+UNet, hybrid+UNet, and the PCA-only method. All UNet-based methods retain the large-scale morphology of the underlying HI distribution, while the PCA-only result suffers from severe suppression and loss of small-scale structures. Notably, the hybrid method produces the most visually accurate reconstruction, recovering both the peak amplitudes and the diffuse structures with high fidelity.

\begin{figure*}
    \centering
    \includegraphics[width=1\linewidth]{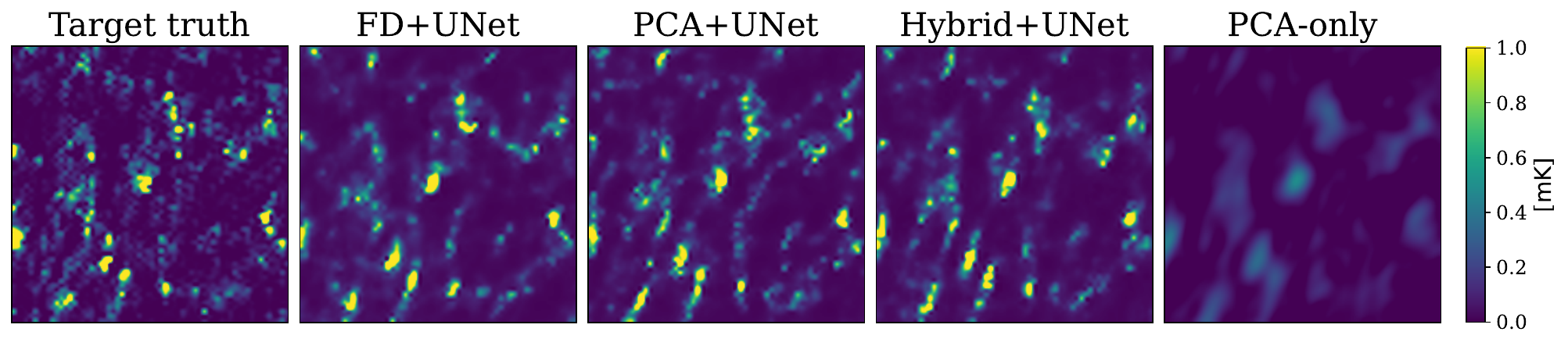}
    \caption{Same as Fig.~\ref{fig:slice_nobeam}, but with cosine beam convolution applied. The results from the hybrid method are also included for comparison.}
    \label{fig:slice_cosine}
\end{figure*}
\begin{figure*}
    \centering
    \includegraphics[width=1\linewidth]{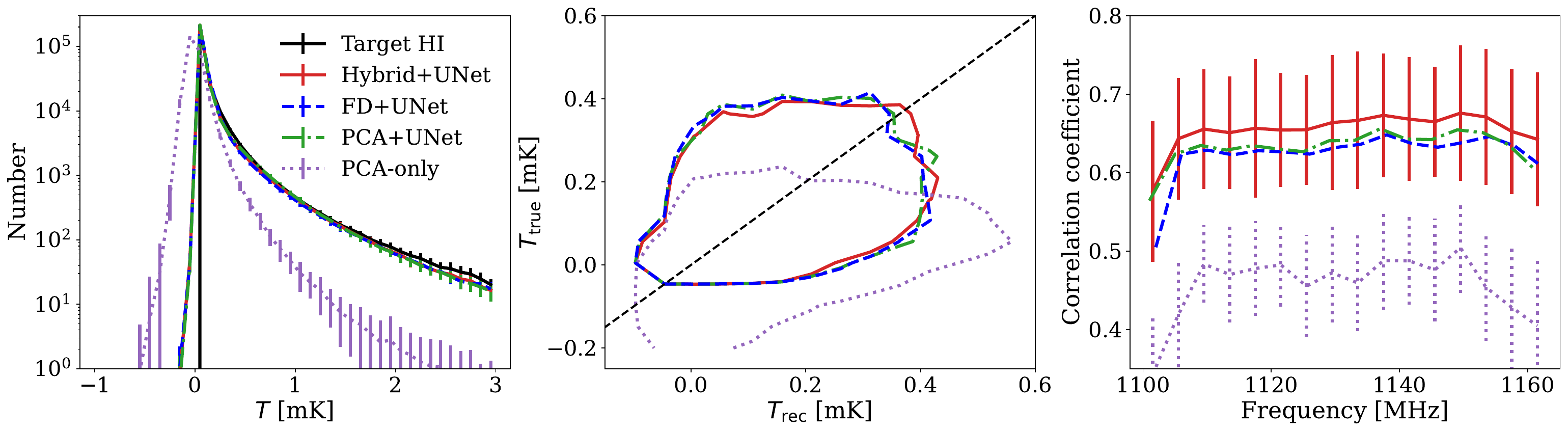}
    \caption{Same as Fig.~\ref{fig:pixel_nobeam}, but with cosine beam convolution applied. The results from the hybrid method are also included for comparison. For clarity, error bars are shown only for the hybrid and PCA-only cases in the right panel.}
    \label{fig:pixel_cosine}
\end{figure*}
\begin{figure*}
    \centering
    \includegraphics[width=1\linewidth]{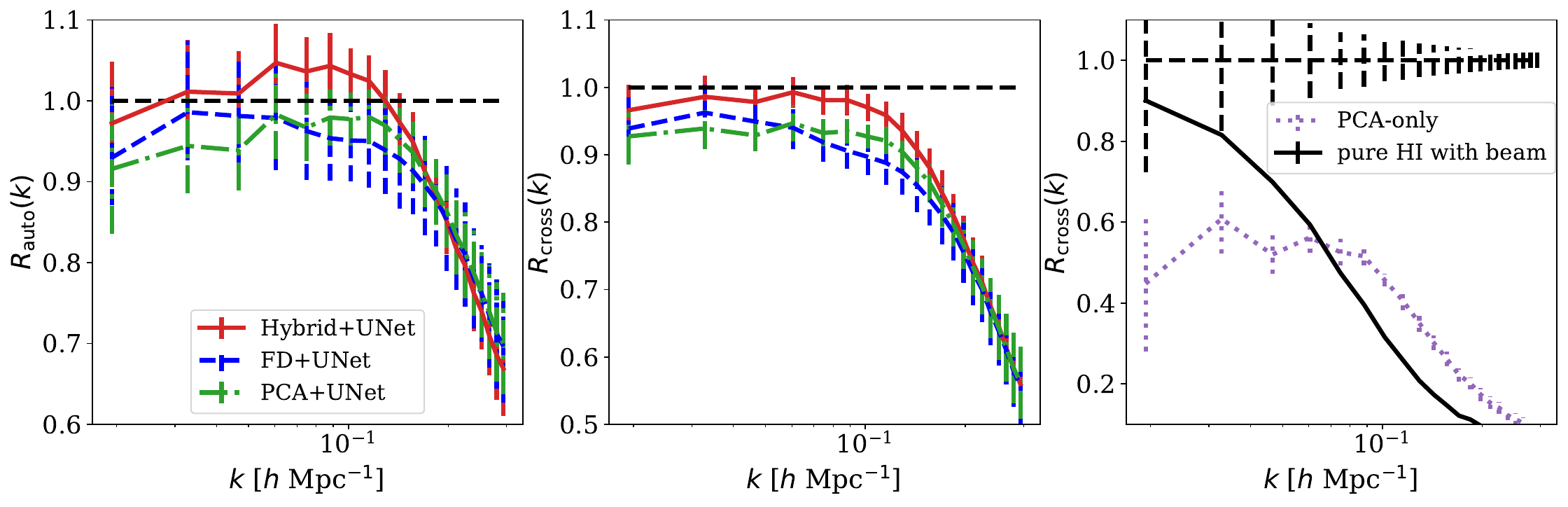}
    \caption{Same as Fig.~\ref{fig:pk1d_gauss}, but with cosine beam convolution applied.}
    \label{fig:pk1d_cosine}
\end{figure*}
\begin{figure*}
    \centering
    \includegraphics[width=1\linewidth]{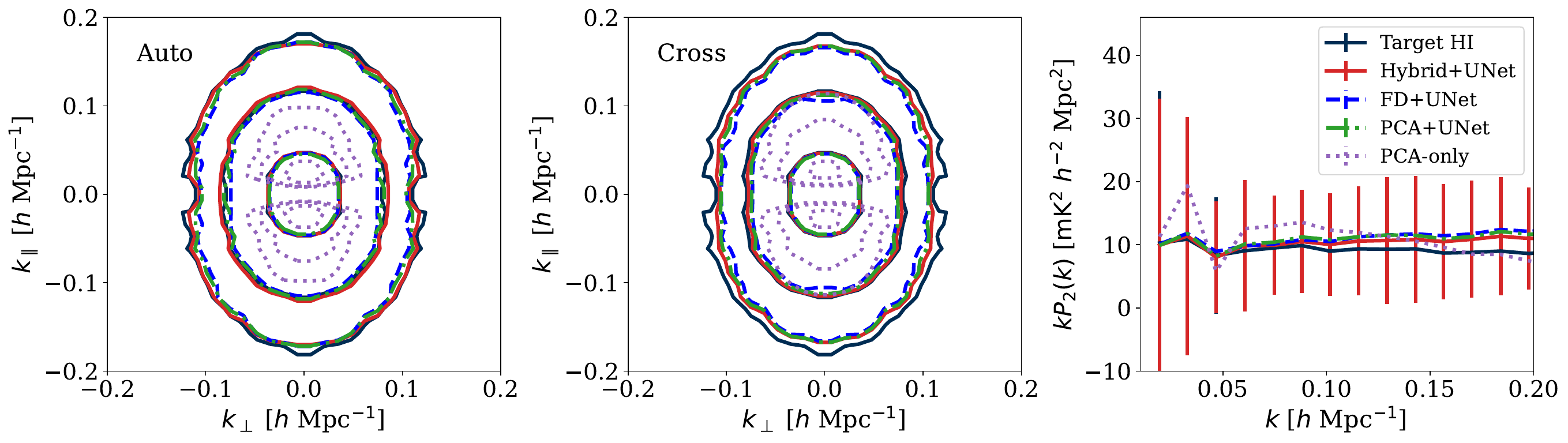}
    \caption{Same as Fig.~\ref{fig:pk2d_nobeam}, but with cosine beam convolution applied. The results from the hybrid method are also included for comparison. For clarity, error bars are shown only for the hybrid case in the right panel.}
    \label{fig:pk2d_cosine}
\end{figure*}

Fig.~\ref{fig:pixel_gauss} presents the temperature distributions (left panel), pixel-to-pixel correlation (middle panel), and correlation coefficient (left panel). All UNet-based methods reproduce the overall shape of the temperature distribution, with only minor differences visible among them. In the pixel-to-pixel correlation, FD+UNet shows a slightly larger spread at low temperatures, suggesting better retention of faint diffuse components, while PCA+UNet more closely follows the diagonal at higher temperatures, indicating improved recovery of strong peaks. This trend is consistent with Fig.~\ref{fig:cross_preprocess}. Frequency differencing preserves large-scale diffuse emission but introduces distortions at high amplitudes, whereas PCA suppresses some large-scale modes yet retains compact bright structures more accurately. The hybrid method combines these strengths and achieves the tightest overall correlation with the ground truth. The correlation-coefficient curves show that all UNet-based models achieve high correlations across the band, with the hybrid method consistently reaching the largest values, while FD+UNet and PCA+UNet remain slightly lower but still closely follow the ground truth. In contrast, the PCA-only reconstruction shows substantially reduced amplitudes and lower correlations, highlighting its limited ability to recover the HI signal in the presence of Gaussian beam smoothing.

Fig.~\ref{fig:pk1d_gauss} presents the auto- and cross-power-spectrum ratios under Gaussian beam convolution. At large scales ($k<0.1 \hmpc$), all UNet-based methods recover power ratios close to unity, indicating that the large-scale structures are well reconstructed. Among them, the hybrid model achieves the best cross-correlation consistency by combining the large-scale mode preservation of frequency differencing with the effective foreground suppression of PCA, thereby reducing scale-dependent deviations. At the same time, a mild positive bias is visible in the reconstructed autopower spectrum over part of the scale range, and this effect is most pronounced for the hybrid method. This behavior can be attributed to the joint use of the two input channels: the FD and PCA residuals contain complementary but partially overlapping signal information, and their combination can reinforce common HI-like modes in the reconstruction, leading to a modest overestimation of the amplitude. Since a weaker positive bias is also present in the FD+UNet result, we do not attribute this effect exclusively to the hybrid combination. Overall, the results indicate that the hybrid method provides improved structural consistency with the true signal while exhibiting a mild residual amplitude bias. At smaller scales ($k>0.1 \hmpc$), all methods show a pronounced decline in both the auto- and cross-power ratios relative to Fig.~\ref{fig:pk1d_nobeam}, primarily reflecting the loss of small-scale power caused by the beam smoothing of fine structures.

The third panel of Fig.~\ref{fig:pk1d_gauss} provides additional context by including the pure HI field convolved with a Gaussian beam and the PCA-only reconstruction as reference cases. The beam-convolved HI serves as the expected baseline illustrating the inherent loss of small-scale power due to instrumental effects, while the PCA result represents the performance of a conventional linear method. Compared to these references, all UNet-based models achieve significantly higher cross-correlation across scales, demonstrating their ability to partially compensate for the beam-induced suppression and recover finer structural information beyond what PCA alone can achieve.

Fig.~\ref{fig:pk2d_gauss} further examines the recovery of RSD information under Gaussian beam convolution. Both FD+UNet and PCA+UNet show good agreement with the target $P(k_\perp, k_\parallel)$, with the anisotropic Kaiser features clearly preserved. The quadrupole $P_2(k)$ remains consistent with the target, confirming that the RSD signal is robustly recovered despite beam smoothing. In contrast, the PCA-only result exhibits an atypical shape in $P(k_\perp, k_\parallel)$ and a biased amplitude in $P_2(k)$, indicating residual contamination.

Overall, these results demonstrate that Gaussian beam smoothing suppresses fine-scale features, but UNet-based reconstructions substantially alleviate this effect. All three methods clearly outperform the PCA-only method, and the hybrid method delivers the most consistent results, particularly at large scales. This highlights the advantage of combining complementary preprocessing strategies when restoring the 21-cm signal under instrumental smoothing.

\begin{figure*}
    \centering
    \includegraphics[width=0.9\linewidth]{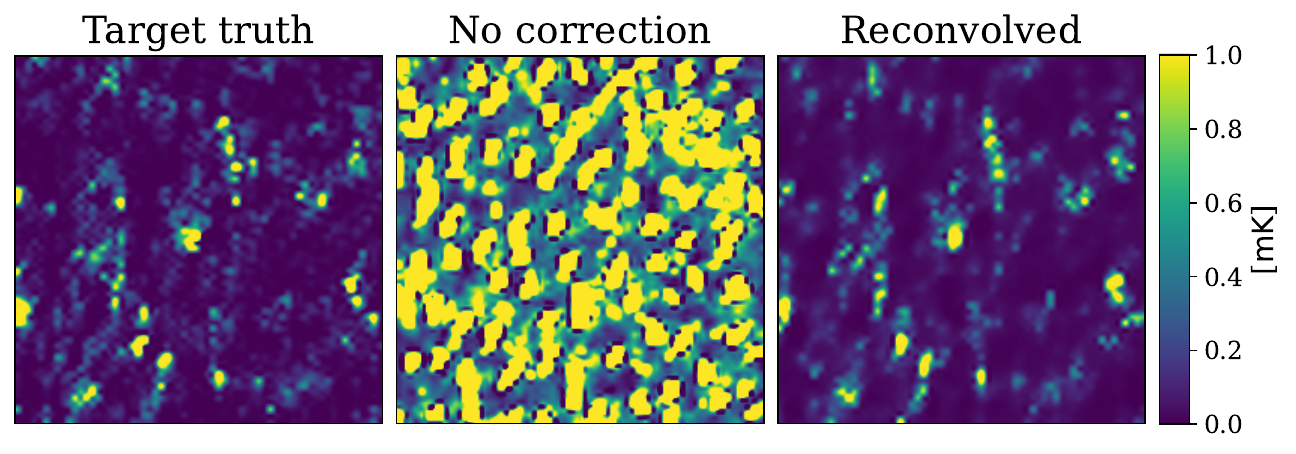}
    \caption{Reconstruction results for beam-mismatch test A. The network is trained with a Gaussian beam model and applied to testing data with a perturbed Gaussian beam. From left to right: target HI map, reconstruction obtained by directly applying the trained model without correction, and reconstruction after applying an additional Gaussian beam reconvolution to the testing data.}
    \label{fig:gaus_smooth}
\end{figure*}
\begin{figure*}
    \centering
    \includegraphics[width=0.9\linewidth]{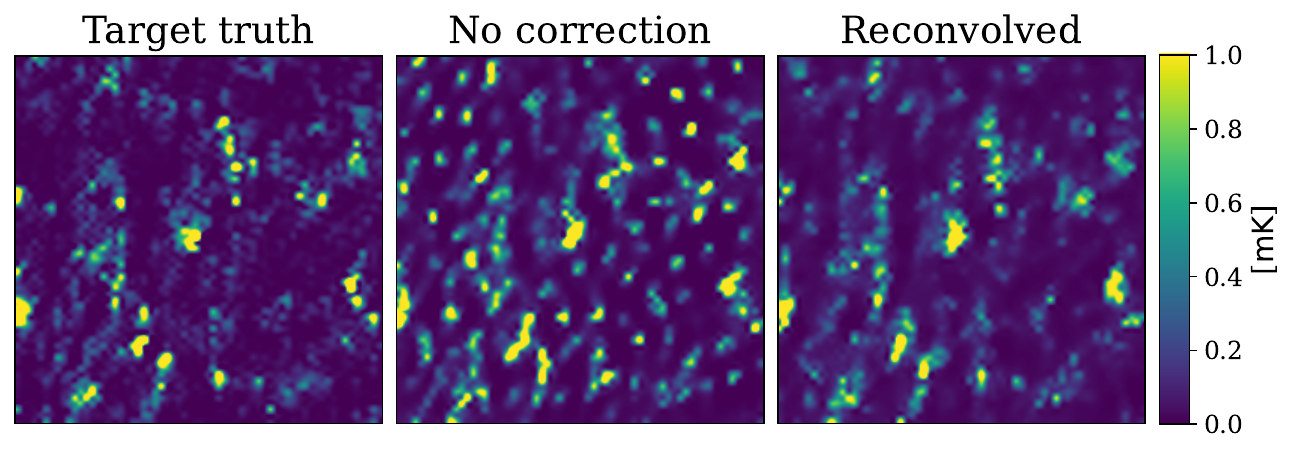}
    \caption{Same as 
    Fig.~\ref{fig:gaus_smooth}, but for beam-mismatch test B, where the network is trained with a Gaussian beam model and applied to testing data with a cosine beam.}
    \label{fig:cos_smooth}
\end{figure*}
\subsection{HI + foreground with cosine beam convolution}

We next assess the reconstruction performance under the more realistic cosine beam model, which introduces frequency-dependent mainlobe narrowing and sidelobe structures that more closely mimic actual instrumental effects. Fig.~\ref{fig:slice_cosine} shows 2D slices of the target HI field and the reconstructed maps under cosine beam convolution. Similar to the Gaussian beam case, all three UNet-based methods successfully recover the large-scale morphology of the HI distribution, whereas the PCA-only result remains strongly suppressed and misses much of the small-scale structure. Among the UNet reconstructions, the hybrid method most closely matches the target map, capturing both bright peaks and diffuse regions with higher fidelity. FD+UNet and PCA+UNet also yield visually consistent reconstructions, though with slightly greater structural differences relative to the target.

Fig.~\ref{fig:pixel_cosine} presents the temperature distributions (left panel), pixel-to-pixel correlations (middle panel), and correlation coefficient (right panel) under cosine beam convolution. Compared with the no-beam (Fig.~\ref{fig:pixel_nobeam}) and Gaussian beam (Fig.~\ref{fig:pixel_gauss}) cases, all methods show increased scatter and stronger departures from the true statistics, reflecting the stronger frequency dependence and sidelobe complexity of the cosine beam. Nevertheless, all three UNet-based methods continue to recover the overall brightness distribution and maintain correlations with the target field, whereas the PCA-only method remains severely biased and strongly suppressed. Among the UNet approaches, the hybrid method achieves the highest correlation coefficients across the band, indicating a modest but persistent advantage under these more challenging beam conditions.

Fig.~\ref{fig:pk1d_cosine} shows the auto- and cross-power-spectrum ratios under cosine beam convolution. Compared with the no-beam and Gaussian beam cases, all three UNet-based methods experience stronger suppression, which is already visible at large scales ($k<\hmpc$) and increases toward higher $k$. At $k < 0.1 \hmpc$, both FD+UNet and PCA+UNet underestimate the cross-power by approximately 5\%–8\%, with the deficit exceeding 20\% around $k\simeq0.2 \hmpc$. In contrast, the hybrid model remains statistically consistent with unity within the $1\sigma$ level across large scales and exhibits relatively smaller deviations at smaller scales, indicating a more accurate recovery of the true power spectrum. This again demonstrates that the hybrid approach most effectively alleviates the large-scale suppression caused by the frequency-dependent sidelobes of the cosine beam. At the same time, the hybrid method exhibits a mild positive bias in the autopower spectrum over part of the scale range, possibly due to the slight reinforcement of overlapping HI signal components when combining the two inputs, as also seen in Fig.~\ref{fig:pk1d_gauss}. The third panel includes the pure HI field convolved with the cosine beam and the PCA-only reconstruction for reference. The strong suppression in the beam-convolved HI case illustrates the intrinsic loss of small-scale power, while the UNet-based models recover much of this loss and achieve noticeably better correlations than the PCA result.

Fig.~\ref{fig:pk2d_cosine} shows the reconstructed 2D power spectra under cosine beam convolution. The overall anisotropic structure in $P(k_\perp,k_\parallel)$ is well preserved by all UNet-based methods, with only a moderate suppression of power compared to the Gaussian case. The hybrid provides the closest agreement with the target, while FD+UNet and PCA+UNet show comparable performance. The quadrupole $P_2(k)$ is also broadly consistent with the target on large scales, indicating that the RSD signal remains robust under this more complex beam model. In contrast, the PCA-only method exhibits clear distortions in both the 2D power spectra and the quadrupole amplitude.

Therefore, the UNet-based methods remain effective in recovering the main features of the HI field under the cosine beam. Notably, the hybrid method shows a clear advantage, achieving much closer agreement with the true power spectra, particularly at large scales. These results highlight the value of combining complementary preprocessing strategies to mitigate the impact of realistic beam effects.

\begin{figure*}
    \centering
    \includegraphics[width=1\linewidth]{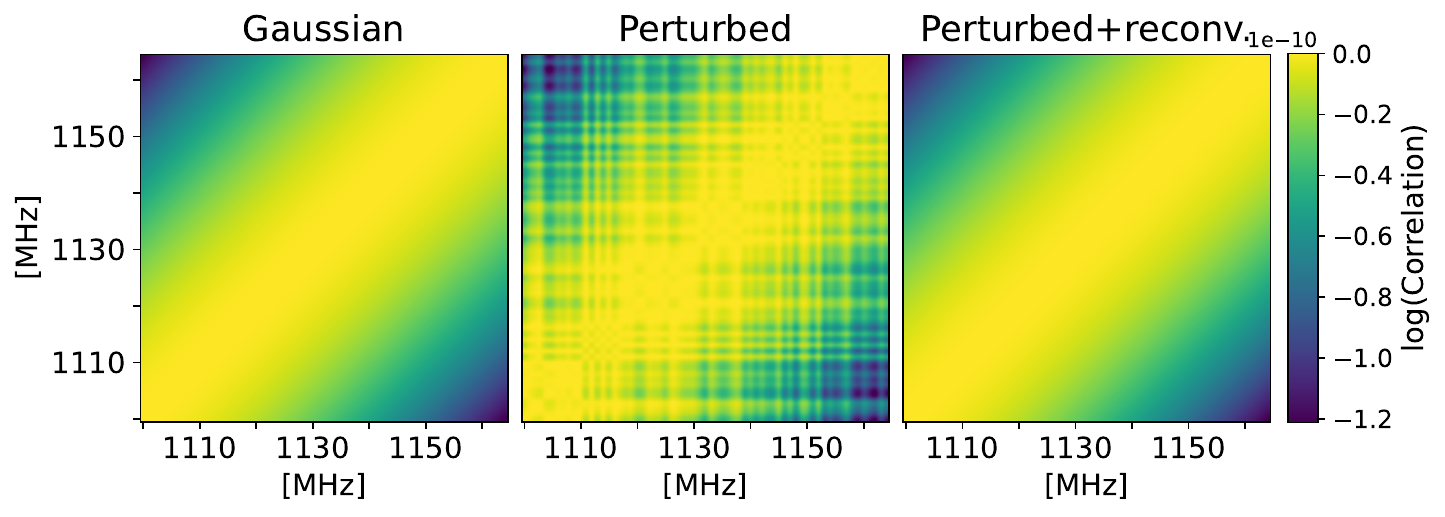}
    \caption{Frequency-frequency covariance matrices of the foreground maps. From left to right: Gaussian beam case, perturbed beam case, and perturbed beam case after additional Gaussian beam reconvolution.}
    \label{fig:corrcoef}
\end{figure*}

\subsection{Performance under beam-model mismatch}\label{sec:mismatch}

All results presented above assume that the beam model used to generate the training set is identical to that used in the testing data. In realistic observations, however, the beam response is not perfectly known. Uncertainties in the beam width, chromatic evolution, or angular profile can therefore introduce a mismatch between the assumed beam model used for training and the true beam affecting the observed maps. Such a mismatch constitutes an out-of-sample effect for the network and may bias the recovered HI signal (e.g., \cite{2024MNRAS.532.2615C}).

To assess the robustness of the proposed method against this effect, we perform two additional beam-mismatch tests in which the training and testing beam models are intentionally made inconsistent:
\begin{itemize}
    \item test A: the testing samples are generated using a perturbed Gaussian-like beam,  in which the FWHM is subject to a stochastic perturbation relative to the Gaussian beam used in training (see Eq.~(\ref{eq:pertbeam}) and the green dashed line in Fig.~\ref{fig:beam_profile}). This case is designed to mimic moderate beam-model uncertainty, such as calibration residuals or imperfect beam fitting within the same beam family.
    \item test B: the testing samples adopt the cosine beam, while the network is still trained using the Gaussian beam model. This represents a more severe mismatch, in which the true beam contains sidelobe and chromatic structures that are absent from the training set.
\end{itemize}

Fig.~\ref{fig:gaus_smooth} presents the reconstruction results for test A. The middle panel shows the result obtained by directly applying the trained model to the mismatched testing sample without any correction. In this case, the reconstructed map is strongly degraded, showing spurious high-intensity regions and poor agreement with the target field. A similar behavior is also observed in test B, as shown in the middle panel of Fig.~\ref{fig:cos_smooth}. These results indicate that the network is not robust to beam-model mismatch when applied directly to out-of-sample data.

The degradation mainly arises because beam mismatch modifies the 
frequency-dependent foreground response and thus distorts its interfrequency correlation feature. Since our pipeline exploits foreground spectral smoothness through both frequency-differencing and PCA, the mismatch-induced fluctuations reduce the effectiveness of foreground removal. This is illustrated by the covariance matrices in Fig.~\ref{fig:corrcoef}, where the nominal Gaussian beam case (left panel) exhibits a smooth and coherent frequency covariance, while the perturbed beam case (middle panel) shows clear striping and nonsmooth features.

\begin{figure*}
    \centering
    \includegraphics[width=1\linewidth]{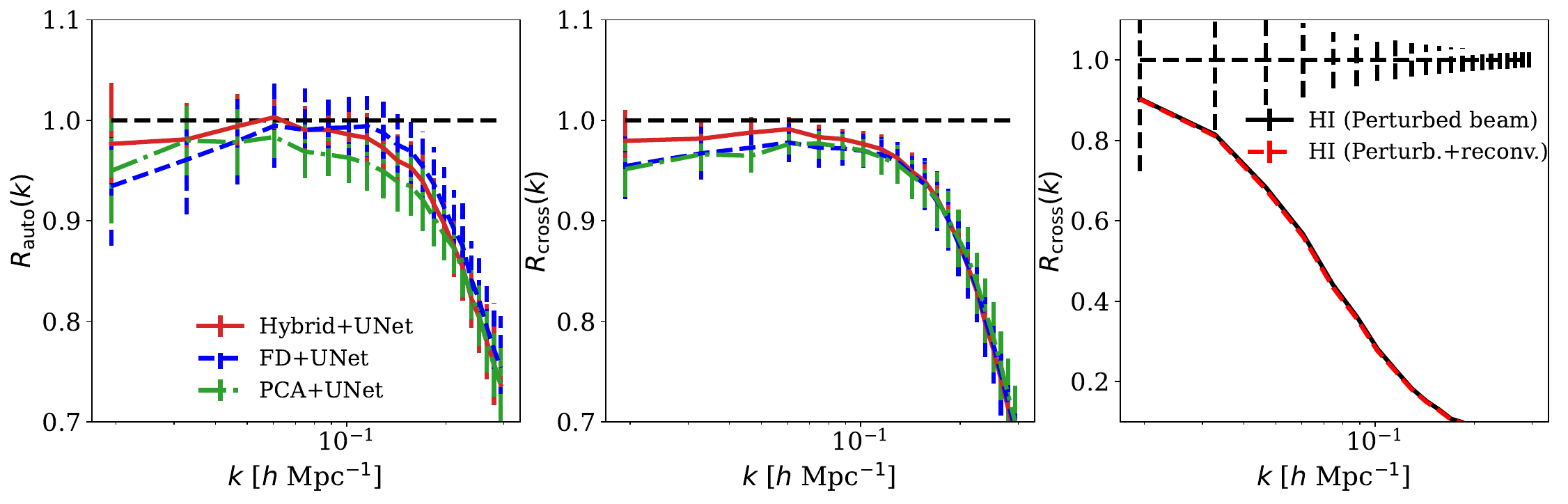}
    \caption{Power-spectrum performance for beam-mismatch test A after applying the additional Gaussian beam reconvolution, with the effective beam FWHM increased by 1\%. The left and middle panels show the reconstructed autospectrum ratio $R_{\rm auto}(k)$ and cross-correlation coefficient $R_{\rm cross}(k)$, respectively, for different foreground-removal pipelines. To illustrate the effect of reconvolution on the power spectrum, the right panel compares the pure HI results with the perturbed beam before (black solid) and after the additional reconvolution (red dashed).}
    \label{fig:Pkgas_smooth}
\end{figure*}
\begin{figure*}
    \centering
    \includegraphics[width=1\linewidth]{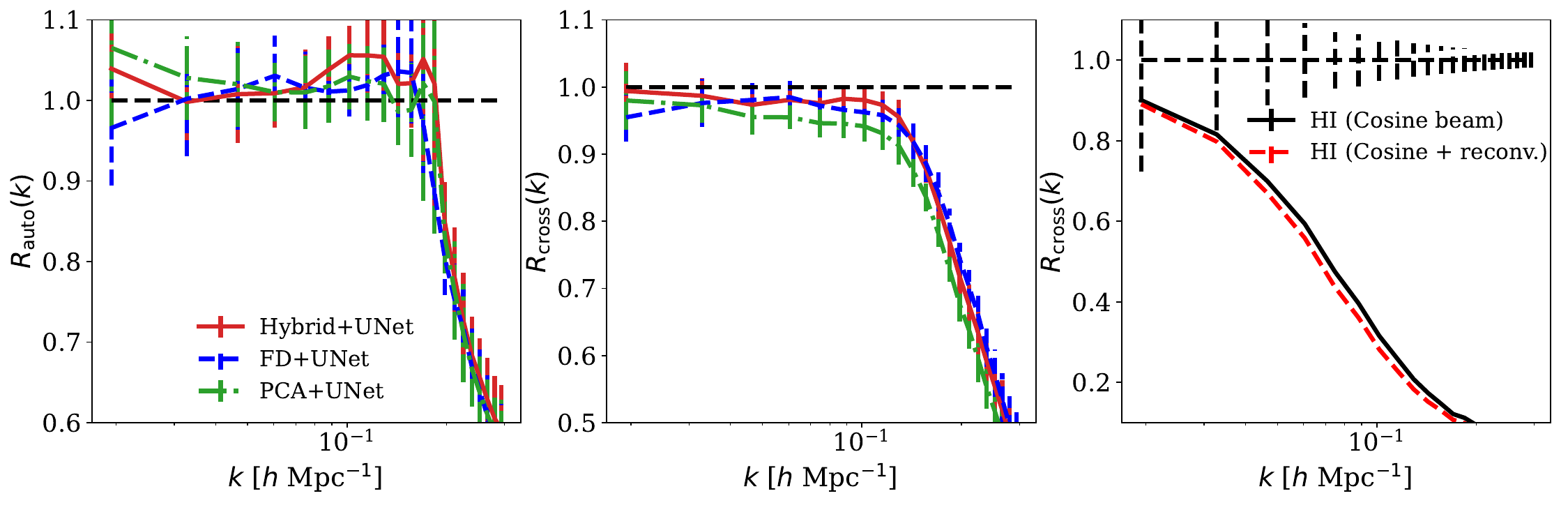}
    \caption{Same as Fig.~\ref{fig:Pkgas_smooth}, but for beam-mismatch test B after applying the additional Gaussian beam reconvolution, with the effective beam FWHM increased by 5\%.}
    \label{fig:Pkcos_smooth}
\end{figure*}

To alleviate this effect, we apply an additional Gaussian beam reconvolution step to the testing maps before feeding them into the trained network. This operation is performed only on the 2D sky projection. The Gaussian kernel is chosen such that the final effective beam FWHM is increased by 1\% and 5\% relative to the original beam for test A and test B, respectively. This step is not intended to exactly reproduce the beam used in training. Instead, it reduces artificial frequency-dependent fluctuations and helps restore the intrinsic spectral smoothness of the foregrounds. As shown in Fig.~\ref{fig:corrcoef}, the covariance matrix (right panel) after smoothing becomes much closer to that of the nominal Gaussian beam case, indicating that the smooth interfrequency correlation structure is largely recovered. Consistently, the right panels of Figs.~\ref{fig:gaus_smooth} and \ref{fig:cos_smooth} show visibly improved coherence of the large-scale structures after reconvolution.

Figs.~\ref{fig:Pkgas_smooth} and \ref{fig:Pkcos_smooth} show the power-spectrum performance for test A and test B, respectively, after applying the additional beam reconvolution. In both cases, the reconstructed autospectrum ratio $R_{\rm auto}(k)$ and cross-correlation coefficient $R_{\rm cross}(k)$ remain close to unity on large scales. This shows that the method can still recover the HI signal robustly even when the beam model is not exactly known and the testing data differ from the training assumption. The overall performance remains comparable to the beam-matched baseline (shown in Figs.~\ref{fig:pk1d_gauss} and \ref{fig:pk1d_cosine}), with the hybrid approach continuing to provide the most stable reconstruction on large scales. Interestingly, the test B reconstruction with beam reconvolution can yield slightly higher cross-correlation than the matched cosine beam case. This is because the additional smoothing suppresses small-scale fluctuations and restores spectral coherence, which improves phase consistency between the reconstructed and true fields. As a result, the cross-correlation is enhanced, even though some small-scale power is reduced.

Moreover, the additional reconvolution does not introduce a substantial large-scale bias in the HI power spectrum. The right panels of Figs.~\ref{fig:Pkgas_smooth} and \ref{fig:Pkcos_smooth} compare the pure HI results with and without this operation. For the 1\% case, the two curves are nearly identical on large scales, with only mild suppression at high $k$. For the 5\% case, a more noticeable suppression extends to a wider range of scales, as expected for a broader effective beam. This is the expected effect of beam smoothing, with power suppression increasing toward smaller scales and only minor impact on large scales.

\section{Summary and conclusion}\label{sec:summary}

In this work, we developed a deep-learning framework for mitigating foreground contamination and instrumental beam effects in 21-cm intensity mapping. The method is built on a UNet architecture and is tested using three alternative preprocessing strategies: FD, PCA, and a two-channel hybrid combination of the two. We tested the approaches using simulated maps with realistic foregrounds and evaluated their performance under Gaussian beam convolution, a frequency-dependent cosine beam model, and additional beam-mismatch scenarios. These setups represent increasingly realistic observational conditions.

The reconstructed maps exhibit clear differences between the deep-learning approaches and the PCA-only baseline. 
The PCA-only method removes spectrally smooth components but consequently suppresses a significant fraction of the cosmological signal when beam convolution is included. This leads to reduced amplitudes and noticeably lower correlation with the true signal. In comparison, the UNet-based reconstructions retain substantially more of the underlying fluctuations, indicating that the network can mitigate the residual foreground contamination and partially correct the distortions produced by the beam.

Our analysis of the UNet-based reconstructions further shows that the preprocessing strategies contribute complementary strengths. Frequency differencing preserves large-scale modes but introduces artificial structures in regions of high brightness temperature, whereas PCA recovers most bright small-scale features but suppresses part of the large-scale signal. These observations motivate the hybrid framework, which combines both inputs through a two-channel architecture to capture the advantages of each preprocessing technique.

Under matched beam conditions, the hybrid preprocessing approach delivers the strongest performance, particularly in the presence of the cosine beam. The cross-correlation power spectrum remains statistically consistent with unity at the $1\sigma$ level on large scales ($k<0.1\,h\,\mathrm{Mpc}^{-1}$), with only modest deviations at higher $k$. In contrast, the FD+UNet and PCA+UNet methods show clear suppression, underestimating the cross-power spectrum by roughly $5\%$–$8\%$ at $k<0.1\,h\,\mathrm{Mpc}^{-1}$ and by more than $20\%$ at $k\simeq0.2\,h\,\mathrm{Mpc}^{-1}$.

Under beam-mismatch conditions, the reconstruction becomes sensitive to inconsistencies between the training and testing beam models, as these distort the interfrequency foreground correlations. However, by applying an additional Gaussian beam reconvolution to the testing maps, the spectral smoothness is partially restored, leading to a substantial improvement in reconstruction quality. The recovered cross-correlation power spectrum returns to a level comparable to the matched-beam case on large scales, demonstrating that the method can still recover the HI signal robustly even when the beam model is not exactly known and the testing data differ from the training assumption.

We note that the current analysis does not imply full robustness to arbitrary or unknown systematics. The beam-mismatch tests presented here should be interpreted as an initial step toward quantifying the impact of out-of-sample effects, rather than a complete characterization of model generalization. While the additional reconvolution can effectively mitigate moderate beam inconsistencies, more complex instrumental uncertainties or foreground modeling errors may introduce residual biases that are not fully captured in the present framework. A more systematic exploration of such effects, including a broader range of instrumental and astrophysical uncertainties, will be essential for establishing the reliability of deep-learning-based reconstruction in realistic survey conditions.

Overall, our findings show that deep learning, particularly when coupled with hybrid preprocessing, provides a powerful and resilient strategy for mitigating both foregrounds and beam effects in 21-cm intensity mapping. The ability to recover unbiased large-scale structure even under frequency-dependent beam distortions is essential for precision cosmological applications such as BAO and RSD measurements in upcoming surveys including MeerKAT and SKA. Future work will incorporate thermal noise, calibration systematics, and real observational data, moving toward a fully end-to-end analysis pipeline for next-generation 21-cm cosmology.

\section*{Acknowledgments}
This work is supported by the National SKA Program of China (2022SKA0110200, 2022SKA0110202, and 2025SKA0160100) and the National Natural Science Foundation of China (No. 12103037, No. 12473097, No. 12203038). F. S. acknowledges the support from the State Key Laboratory of Dark Matter Physics and the Young Data Scientist Program of the China National Astronomical Data Center (No. NADC2025YDS-01). This work is supported by the China Manned Space Project with No. CMS-CSST-2021 (A02, A03, B01), and Guangdong Basic and Applied Basic Research Foundation (2024A1515012309). This research was supported by Zhejiang Provincial Natural Science Foundation of China under Grant No. LY24A030001. This work is supported by Natural Science Basic Research Program of Shaanxi (Program No. 2025JC-YBMS-016). The authors acknowledge the Beijing Super Cloud Center (BSCC) for providing HPC resources that have contributed to the research results reported in this paper.

\section*{Data availability}
The data that support the findings of this article are openly available \cite{wang_2025_17648112}.



\bibliography{apssamp}

\end{document}